\documentclass{article}[11pt]
\usepackage{jheppub}
\usepackage[toc,page]{appendix}
\pdfoutput=1
\usepackage{graphicx}
\usepackage{amsmath,amssymb,mathrsfs}
\usepackage{bbm}
\usepackage{color}
\usepackage{xcolor}
\usepackage{dsfont}
\usepackage{cancel}
\usepackage{dsfont}
\usepackage{epstopdf}
\usepackage{epsfig}
\usepackage{bm}
\usepackage{dcolumn}
\usepackage{hyperref}
\usepackage{enumitem}
\usepackage{multirow}
\usepackage{lineno}
\usepackage{mathtools,slashed}
\usepackage{url}
\usepackage{lineno}
\usepackage{wrapfig}
\usepackage{braket}

\begin{document}


\title{Large-Scale Atom Interferometry for Fundamental Physics}

\abstract{
Atom interferometers measure quantum interference patterns in the wave functions of cold atoms that follow superpositions of different space-time trajectories. These can be sensitive to phase shifts induced by fundamental physics processes such as interactions with ultralight dark matter or the passage of gravitational waves. The capabilities of large-scale atom interferometers are illustrated by their estimated sensitivities to the possible interactions of ultralight dark matter with electrons and photons, and to gravitational waves in the frequency range around 1 Hz, intermediate between the peak sensitivities of the LIGO and LISA experiments. Atom interferometers can probe ultralight scalar couplings with much greater sensitivity than is currently available from probes of the Equivalence Principle. Their sensitivity to mid-frequency gravitational waves may open a window on mergers of masses intermediate between those discovered by the LIGO and Virgo experiments and the supermassive black holes present in the cores of galaxies, as well as fundamental physics processes in the early Universe such as first-order phase transitions and the evolution of networks of cosmic strings.
~~\\
~~\\
~~\\
~~\\
~~\\
~~\\
AION-REPORT/2023-04, KCL-PH-TH/2023-33, CERN-TH-2023-105\\
~~\\
~~\\
~~\\
~~\\
~~\\
~~\\
~~\\
}

\author[1,2]{Oliver~Buchmueller,}
\author[3]{John~Ellis,} 
\author[4]{Ulrich~Schneider}  
\affiliation[1]{Imperial College, Physics Department, Blackett Laboratory, Prince Consort Road, London, SW7 2AZ, UK}
\affiliation[2]{University of Oxford, Department of Physics, Clarendon Laboratory, South Parks Road, Oxford OX1 3PU, UK}
\affiliation[3]{Theoretical Particle Physics and Cosmology Group, Department of Physics, King's College London, Strand, London, WC2R 2LS, UK}
\affiliation[4]{Cavendish Laboratory, University of Cambridge, J.J. Thomson Avenue, Cambridge, CB3 0HE, UK}


\maketitle
\section{Introduction}

Quantum technologies offer new prospects for sensitive and precise measurements
in fundamental physics. Prominent among quantum sensors are atom interferometers, and in this article we review their potential applications
to some of the most challenging topics in fundamental physics, namely the
nature of dark matter and measurements of gravitational waves.

For some 80 years now, astronomers and cosmologists have been accumulating
evidence that the visible matter in the Universe floats in invisible clouds
of dark matter~\cite{Zwicky:1933gu,Zwicky:1937zza,Rubin:1970zza,Bertone:2016nfn} that exert gravitational attraction but can have at most very 
weak electromagnetic and other interactions with the visible matter. There
are strong constraints on the fraction of the dark matter that could be
provided by black holes~\cite{Carr:2021bzv} or other compact astrophysical objects, and the
general expectation is that dark matter is composed of one or more species
of so far undetected neutral particles. There are two general schools of thought about
the properties of these dark matter particles, which might be either
weakly-interactive massive particles (WIMPs) or ultralight bosonic
fields that form coherent waves moving non-relativistically through
the Universe. In the absence of evidence for the production of any WIMPs
at particle colliders (see, e.g., \cite{ATLAS:2022ihe,CMS:2023ktc}) or for their collisions with heavy nuclear targets
deep underground~\cite{LZ:2022ufs}, attention is shifting towards ultralight dark matter
(ULDM) candidates that have very weak interactions with the particles of the
Standard Model (SM) that make up the visible matter in the Universe. 
Atom interferometers enable very precise measurements of atomic properties, so are exquisitely sensitive to the possible interactions of
ULDM fields with SM particles~\cite{Graham:2015ifn}. 

Atom interferometers can also probe the small distortions of space-time
caused by the passage of gravitational waves (GWs)~\cite{Graham:2012sy}. The
discovery of GWs by the LIGO and Virgo laser interferometers~\cite{LIGOScientific:2016aoc} opened
a new window on the Universe through which novel phenomena
could be observed, such as the mergers of black holes (BHs)
and neutron stars (NSs)~\cite{LIGOScientific:2017vwq}. It is known that supermassive BHs
(SMBHs) far more massive than those revealed by LIGO and 
Virgo are present in the cores of galaxies~\cite{EventHorizonTelescope:2019dse,EventHorizonTelescope:2022wkp}, and their
binary systems and mergers are the prime target of the planned LISA space-borne
laser interferometer~\cite{Audley:2017drz}. The existence of intermediate-mass BHs
(IMBHs) is strongly suspected~\cite{Greene:2019vlv}, and measurements of the GWs
emitted during their mergers could cast light on the
assembly of SMBHs. As we discuss in more detail below,
atom interferometers can observe
GWs in the mid-frequency range intermediate between LISA and
LIGO/Virgo, and are hence ideal for addressing this issue.
Atom interferometers can also search for possible fundamental-physics sources of GWs,
such as first-order phase transitions in the early
Universe and networks of cosmic strings~\cite{Badurina:2021rgt}.

In addition to these primary objectives, atom
interferometers  can also provide
unique tests of general relativity, e.g., by probing
Einstein's Equivalence Principle~\cite{Asenbaum:2020era}, 
as well as testing the limits of quantum mechanics~\cite{BILARDELLO2016764} and
probing quantum effects in gravitational fields~\cite{Overstreet:2021hea}.
Beyond making significant contributions to fundamental physics,
atom interferometry can also be used to study the properties of ultracold atoms, such as Bose-Einstein condensates and degenerate Fermi gases.
Moreover, the precision of measurements using atom interferometry offers unprecedented sensitivity and accuracy for a variety of physical quantities, such as acceleration, local gravity, and rotation.
For example, atom interferometry can be used to create highly sensitive inertial sensors for navigation and geophysical exploration. Precision measurements of the local gravitational acceleration are of relevance to earth observation and to detect underground structures~\cite{Stray2022}.

In view of its many potential applications to precision measurements and quantum technology, atom interferometry is a rapidly-developing field of research. Many of the practical challenges involved in the design and operation of atom interferometry are being actively addressed, and much research is underway to explore fully its possibilities.

The outline of this article is the following. We review
the basic principles of atom interferometry in Section~\ref{Nutshell}, and in Section~\ref{Landscape} we
review current projects to scale up atom
interferometers to baselines ranging from
${\cal O}(10)$~m to ${\cal O}(1)$~km on Earth, and potentially
thousands of km in space. We then discuss in more detail
in Section~\ref{Applications} the potential applications 
to fundamental physics that motivate these projects,
principally searches for ULDM and GWs. 

\section{Atom Interferometry in a Nutshell}
\label{Nutshell}



Atom interferometry is a relatively new and exciting field of research that explores and exploits the wave-like properties of atoms in setups analogous to optical interferometers. The study of matter waves and their interference  dates back to the early days of quantum mechanics, but the development of laser cooling and trapping techniques in the 1980s and 1990s greatly expanded the possibilities for this area of research. 

The concept of wave-particle duality was first proposed by Louis de Broglie in the 
1920s~\cite{deBroglie1924}. It stipulates that all particles, including atoms, have wave-like properties and that these can give rise to observable interference patterns, for instance when a beam of particles passes through two closely-spaced slits, in the same spirit as Thomas Young's famous double-slit experiment in optics~\cite{young1804experiments}. This proposal was later experimentally confirmed for electrons in independent experiments by Davisson and Germer~\cite{Davisson1927}  and by  Thomson~\cite{Thomson1927}  in 1927.  Esterman and Stern in 1930 extended these observations to atoms by diffracting a beam of sodium atoms off a surface of NaCl~\cite{Estermann1930}. These early observations suggested that the principle of interferometry could also be extended to matter waves. 


The development of lasers beginning in the 1950s enabled not only ever more precise spectroscopy but also more precise control over the internal states and momenta of atoms. In particular, lasers could be used to cool and trap atoms, creating a new field of research using cold atoms. In 1995, the groups of Eric Cornell and Carl Wieman at the University of Colorado and of Wolfgang Ketterle at the Massachusetts Institute of Technology created for the first time a Bose-Einstein condensate (BEC), a new state of matter in which a group of atoms behaves as a coherent entity~\cite{Anderson:1995gf,Davis95}. The creation of BECs opened new possibilities for the study of matter waves.

The first complete atom interferometers were reported in 1991: Carnal and Mlynek~\cite{Carnal1991} and the Pritchard group~\cite{Keith1991} independently demonstrated atom interferometers using microfabricated diffraction gratings. At the same time, the first atom interferometers based on laser pulses were demonstrated independently in the group of Bord\'e at PTB~\cite{Riehle1991} and by Mark Kasevich and Steven Chu in Stanford~\cite{kasevich91corr}.  For a general introduction and review of atom interferometry and some applications see~\cite{Cronin2009}.





\subsection{Principle}
Atom interferometers are most easily understood by analogy to optical interferometers: by splitting and recombining beams of light, the latter are capable of detecting extremely small differences in the phase of the light waves accrued over the different paths, allowing for the measurement of tiny changes in distance, angular velocity, and other physical quantities. Laser interferometers have revolutionized our ability to make precise measurements in a wide range of fields, from astronomy to nanotechnology, and atom interferometers will expand this range even further. 

\begin{figure}[h]
\centering 
\includegraphics[width=7cm]{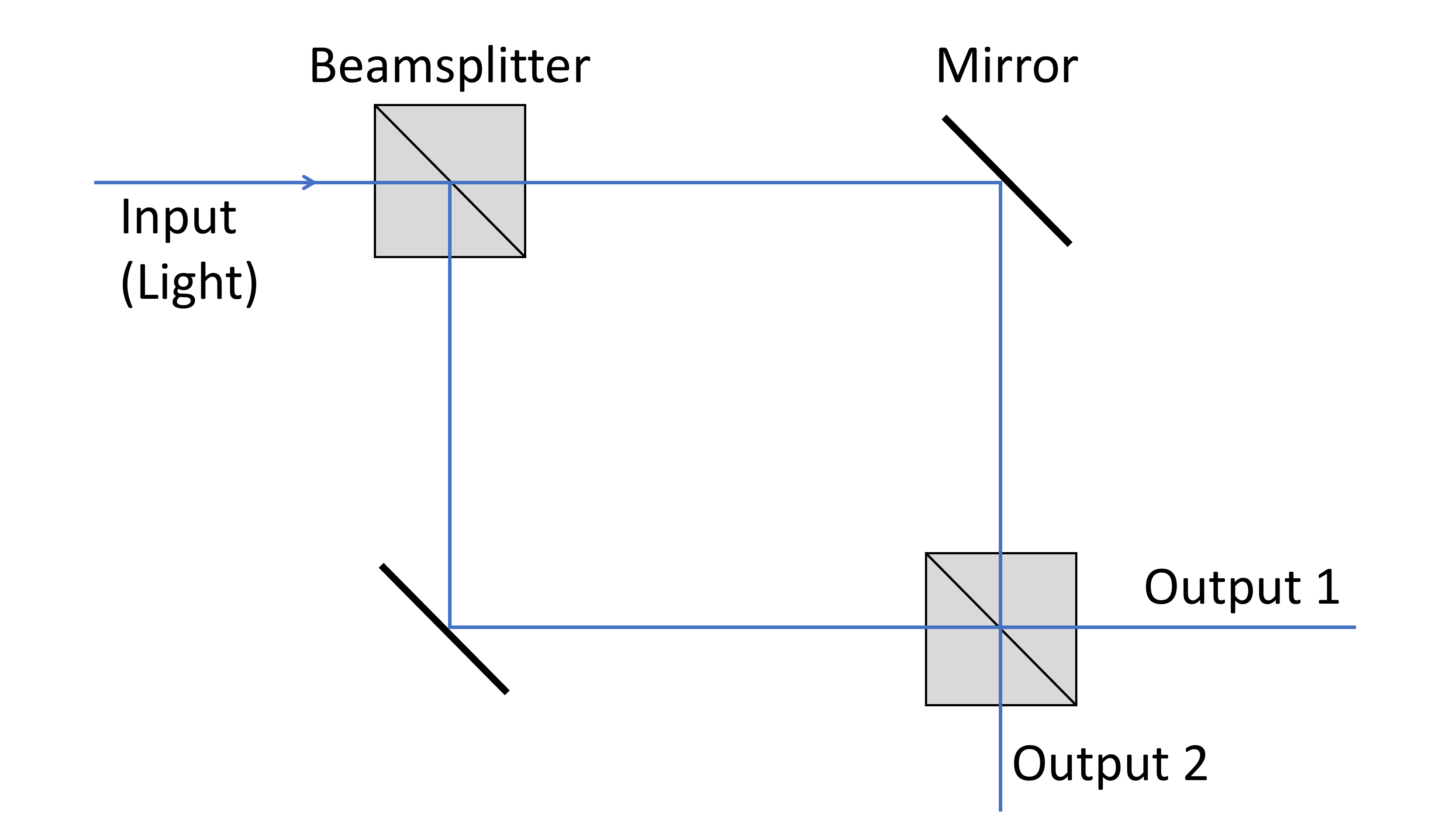}
\includegraphics[width=7cm]{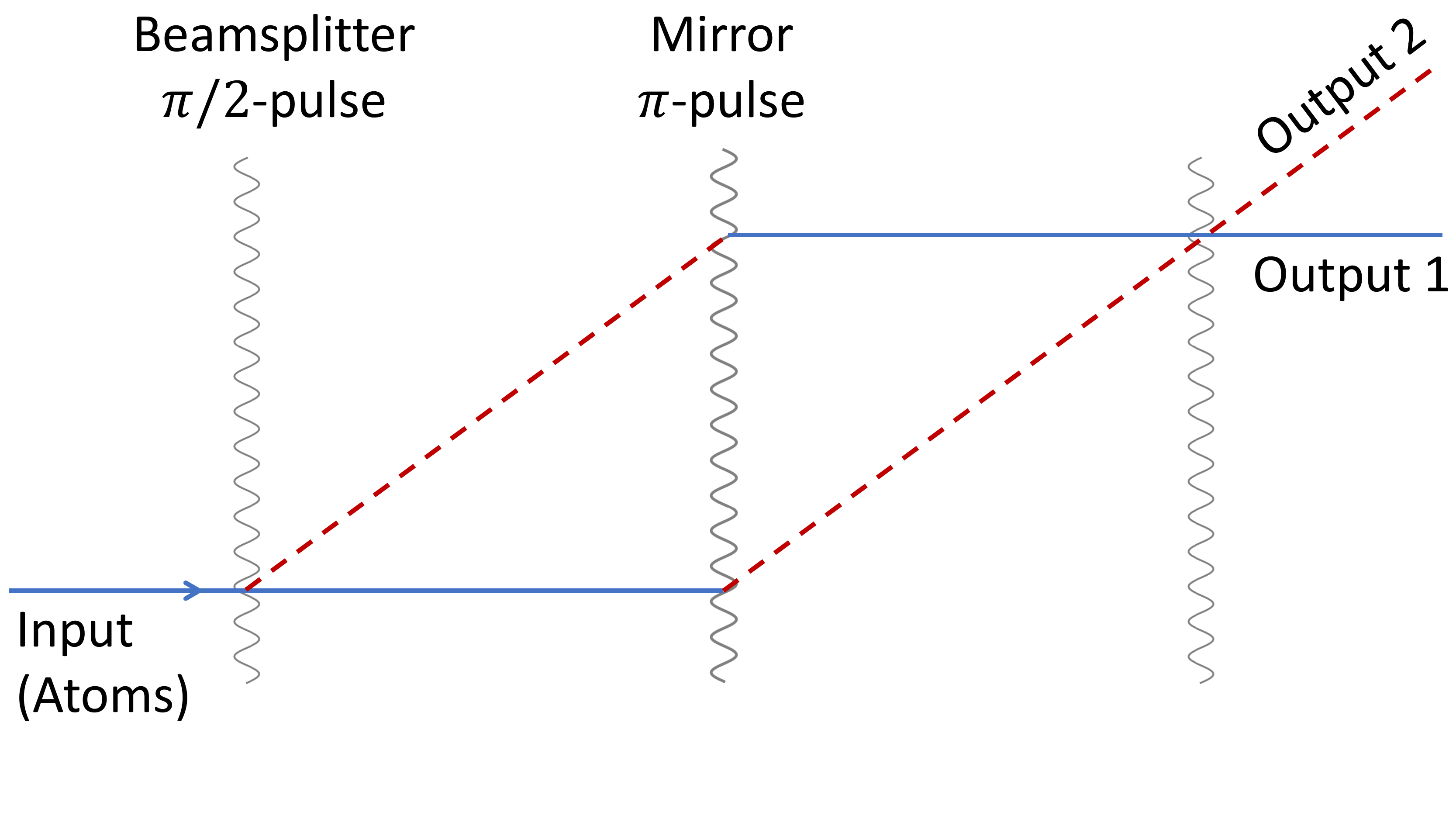}
\caption{\it Left: Conceptual outline of a Mach-Zehnder laser interferometer~\cite{zehnder1891neuer,mach1892ueber}.
Right: Conceptual outline of an analogous atom interferometer. Atoms in the ground state, $\ket{g}$, are represented by solid blue lines, the dashed red lines represent atoms in the excited state, $\ket{e}$, and laser pulses are represented by wavy lines.}
\label{fig:Interferometers}
\end{figure}

Any interferometer will rely on the same basic ingredients, most easily visualized in the Mach-Zehnder configuration~\cite{zehnder1891neuer,mach1892ueber} illustrated in  Fig.~\ref{fig:Interferometers}.
Starting from a coherent source, such as a laser in optical interferometry, a first beamsplitter splits the beam into two parts that travel along different paths, where they can pick up different phases. A mirror is then used to bring the  beams back together so that they can interfere at the final (closing) beamsplitter.  Crucially, the intensities at the two output ports will depend on the differential phase $\Delta\varphi=\phi_1-\phi_2$ picked up along the two arms of the interferometer.

While atom interferometers can be built using microfabricated diffraction gratings, the interferometers discussed in this review rely on atom-light interactions.  They have been made possible by the development of laser technology and quantum optics to levels where the central elements of any interferometer, namely the beamsplitters and mirrors, can be replaced by atom-light interactions, as illustrated in the right panel of Fig.~\ref{fig:Interferometers}.  

\begin{figure}[h]
\centering 
\includegraphics[width=10cm]{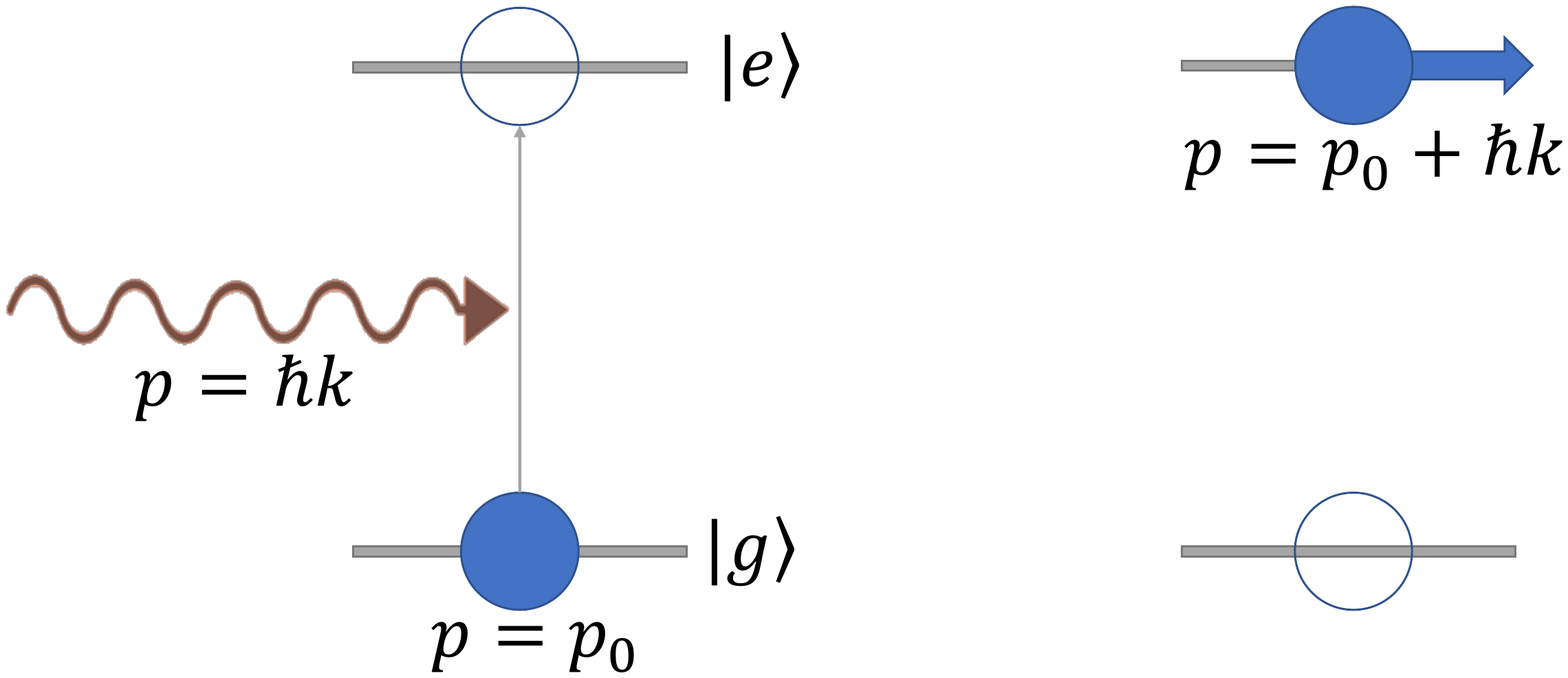}
\vspace{-10mm}
\caption{\it Every photon carries a momentum $p=\hbar k$ and, upon absorption of the photon, its momentum is transferred to the atom. This forms the basis for manipulating atomic momentum with resonant light.}
\label{fig:MomentumKick}
\end{figure}

The laser pulses consist of coherent single-frequency light that is resonant, or close-to-resonant, with the transition between the atomic ground state and a specific excited state ($\ket{g}\leftrightarrow\ket{e}$). As illustrated in Fig.~\ref{fig:MomentumKick} for an atom with initial momentum $p_0$, the absorption (emission) of a photon of wavelength $\lambda$ not only exchanges the atom's internal state  but in the process also changes the atomic momentum by the photon's momentum $\hbar k$ ($-\hbar k$), where $k=2\pi/\lambda$. This recoil kick thereby deflects the atom, as seen in the right panel of Fig.~\ref{fig:Interferometers}.  
By precisely controlling the amplitude and duration of the light pulse, it is possible to implement a $\pi/2-$pulse that can act as a beamsplitter by bringing the atom into an equal superposition of ground and excited states whose momenta differ by $\hbar k$. After this splitting, the two components propagate along their respective paths and may accrue different phase shifts due to, for instance, different magnetic, electric, or gravitational fields.  After a time $T$, a $\pi-$pulse swaps ground and excited states and imparts a further momentum kick of $\pm\hbar k$ that acts as a mirror, so that the paths are eventually recombined. A final $\pi/2-$pulse then acts as the final beamsplitter before the numbers of atoms in the ground and excited states are read out by, e.g., fluorescence imaging. These final atom numbers provide the output ports of this interferometer. 

This $\pi/2-\pi-\pi/2$ or Ramsey pulse sequence is directly analogous to the operation of an atomic clock~\cite{Ludlow2015}. The main difference is that in an interferometer the two parts of the wavefunction become spatially separated and can therefore pick up different spatially-dependent phases. The resulting interference pattern provides information about the physical properties of the system, such as acceleration, rotation, or gravitational field strength. This was demonstrated successfully in 1991 by the group of Bord\'e that observed the Sagnac effect in a rotating apparatus~\cite{Riehle1991}. In the following years, several types of atom interferometers were developed, such as Mach-Zehnder, Ramsey-Bord\'e, and Sagnac interferometers, each with its own strengths and weaknesses~\cite{Cronin2009}. 

In another incarnation of atom interferometry, standing waves of light are used  as a diffraction grating for the atoms~\cite{rasel1995atom} in a similar fashion as crystals can be used to diffract electrons. The standing wave creates a periodic potential that interacts with the atoms, causing them to diffract into two or more paths.  This remains the method of choice when implementing interferometers using ever-larger molecules to test the possible breakdown of quantum mechanics in large-mass systems~\cite{Kialka2022}.

\subsection{Sensitivity}
The achievable sensitivity for an atom interferometer can be decomposed into two parts. The readout sensitivity, that is the precision with which the final phase of the interferometer can be read out, can be assumed to be limited by the shot-noise limit, i.e., to scale as $\propto 1/\sqrt{N}$, where $N$ is the number of atoms. Current interferometers operate in the regime of $10^6$ atoms per second~\cite{Rudolph:2019vcv}, and one area of development for the coming years will be to boost this by several orders of magnitude.
An additional boost to the readout sensitivity will come from squeezing, as discussed below.

The second part of the total sensitivity is the intrinsic sensitivity, i.e., how big a phase shift results from a given strength of the targeted physical effect. It is hence also important to design the interferometer to maximise the phase shifts created by the physics in question while minimising the sensitivity to unwanted systematics. Some important considerations are reviewed briefly below.

\paragraph{Gradiometer}
A fundamental limitation of atom interferometry stems from the phase noise of the interferometer (interrogation) laser used to implement the beamsplitters. As in an atomic clock, the final populations in the output ports depend on the phase difference between the interrogation laser during the final $\pi/2-$pulse and the relative phase between the components of the atomic wave function. Single interferometers are therefore intrinsically susceptible to laser noise. This sensitivity to laser noise forms the basis for atomic clocks, where the resulting error signal is used to stabilize the laser to the atomic transition, giving rise to excellent precision for the time-averaged frequency~\cite{Ludlow2015}. However, it represents a significant limitation for interferometers that aim to study time-dependent effects such as gravitational waves or oscillating dark matter fields.

\begin{figure}
\centering 
\includegraphics[width=14cm]{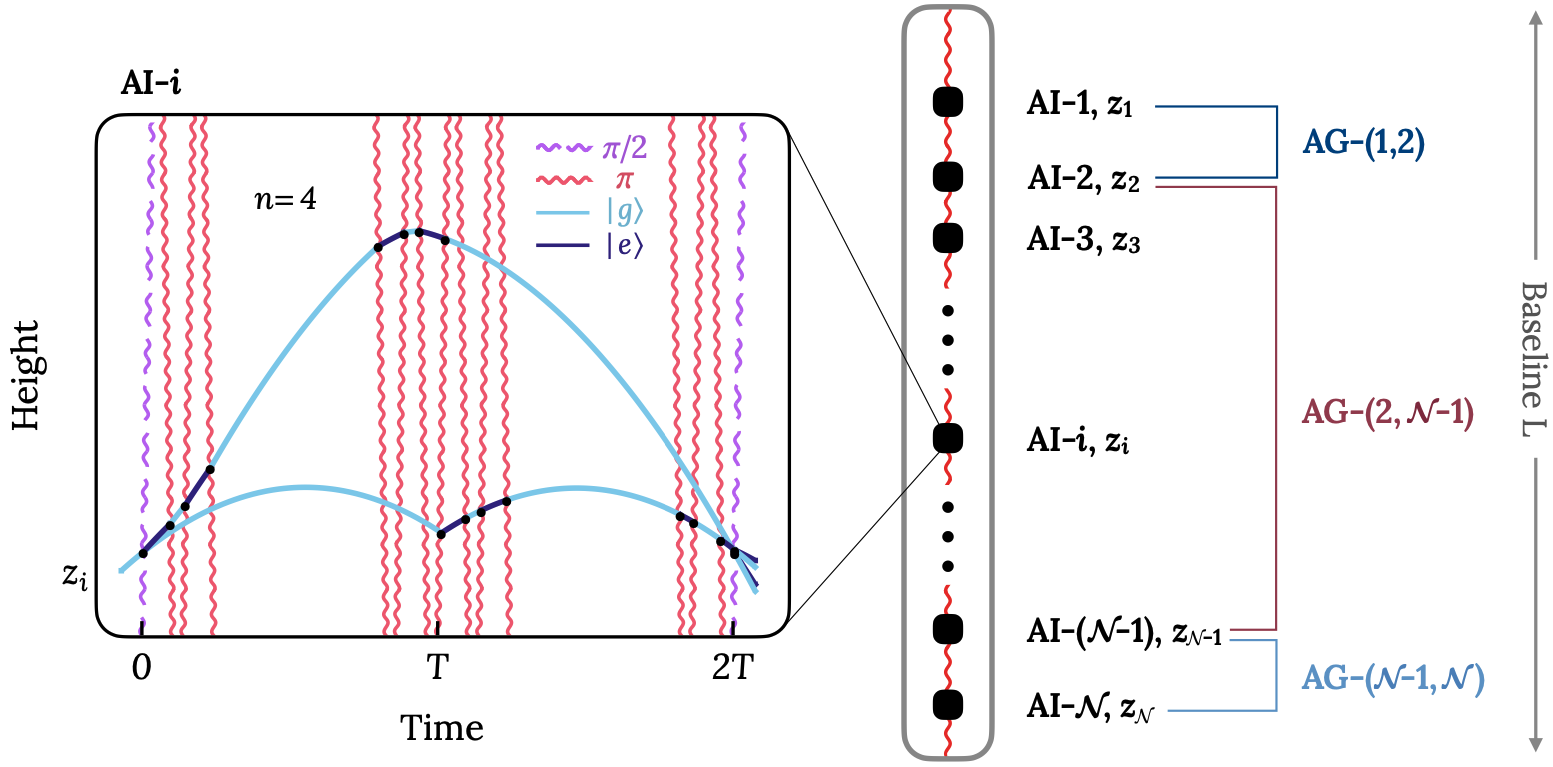}
\vspace{-3mm}
\caption{\it Schematic representation of an gradiometer with multiple atom interferometers. The left panel illustrates the spacetime diagram of one of the atom interferometers with $n = 4$ LMT momentum kicks. The atoms' excited ($|e\rangle$) and ground ($|g\rangle$) states are shown in dark and light blue, respectively, and $\pi/2-$ and $\pi-$pulses are displayed as wavy purple and red lines. The right panel shows how a series of such atom interferometers may be spaced within a single vertical vacuum tube of length $L$, forming multiple atom gradiometers that combine the atom interferometers $i,j$, labelled as AG$-(i,j)$. Figure taken from~\cite{Badurina:2022ngn}.}
\label{fig:Multigradiometer}
\end{figure}

An elegant solution for mitigating this limitation is provided by gradiometer configurations, where two or more interferometers are interrogated by a common laser~\cite{Yu2011},
as illustrated in the right panel of Fig.~\ref{fig:Multigradiometer}. In such a configuration, the laser noise becomes a common mode for all interferometers and hence does not contribute to differential signals, i.e., the differences between the phases measured by the interferometers. Such configurations are by construction not sensitive to common fields, which would induce identical phases in the interferometers, but only to the gradients of, for instance, electric or magnetic fields, and hence are known as gradiometers.
Their sensitivity scales linearly with the separation $\Delta r$ between the interferometers. For terrestrial interferometers this separation is in practice limited to at most the kilometre scale, since the laser needs to travel in vacuum between the interferometers in order to avoid unwanted phase fluctuations from air turbulence or fibre noise. 

\paragraph{Large Momentum Transfer (LMT)}
Large momentum transfers in atom interferometry are analogous to the use of cavities in optical interferometers, where $n$ round trips result in an $n$-times larger overall phase shift. In an LMT pulse sequence, the atoms interact $n$ times with counter-propagating interferometer lasers, as illustrated in the left panel of Fig.~\ref{fig:Multigradiometer}, and thereby acquire a large momentum kick of  $n2\hbar k$. In the process, the atoms get exposed $n$ times to the laser phase, which ultimately increases the sensitivity of the interferometer $n$-fold and thereby enables high-precision measurements. In the last few years LMTs up to 400$\hbar k$ have been demonstrated experimentally~\cite{wilkason2022atom}. 
Future developments will aim at significantly increasing the amount of LMT and will need to overcome the challenge that the interferometer contrast typically decays with increasing LMT due to imperfections in the pulses used and  effects of the finite initial temperature of the atoms, which gives rise to inhomogeneous Doppler shifts. Mitigation strategies will likely include more complex pulse sequences and developments of higher power laser sources.

\begin{figure}
    \centering
    \includegraphics[width=10cm]{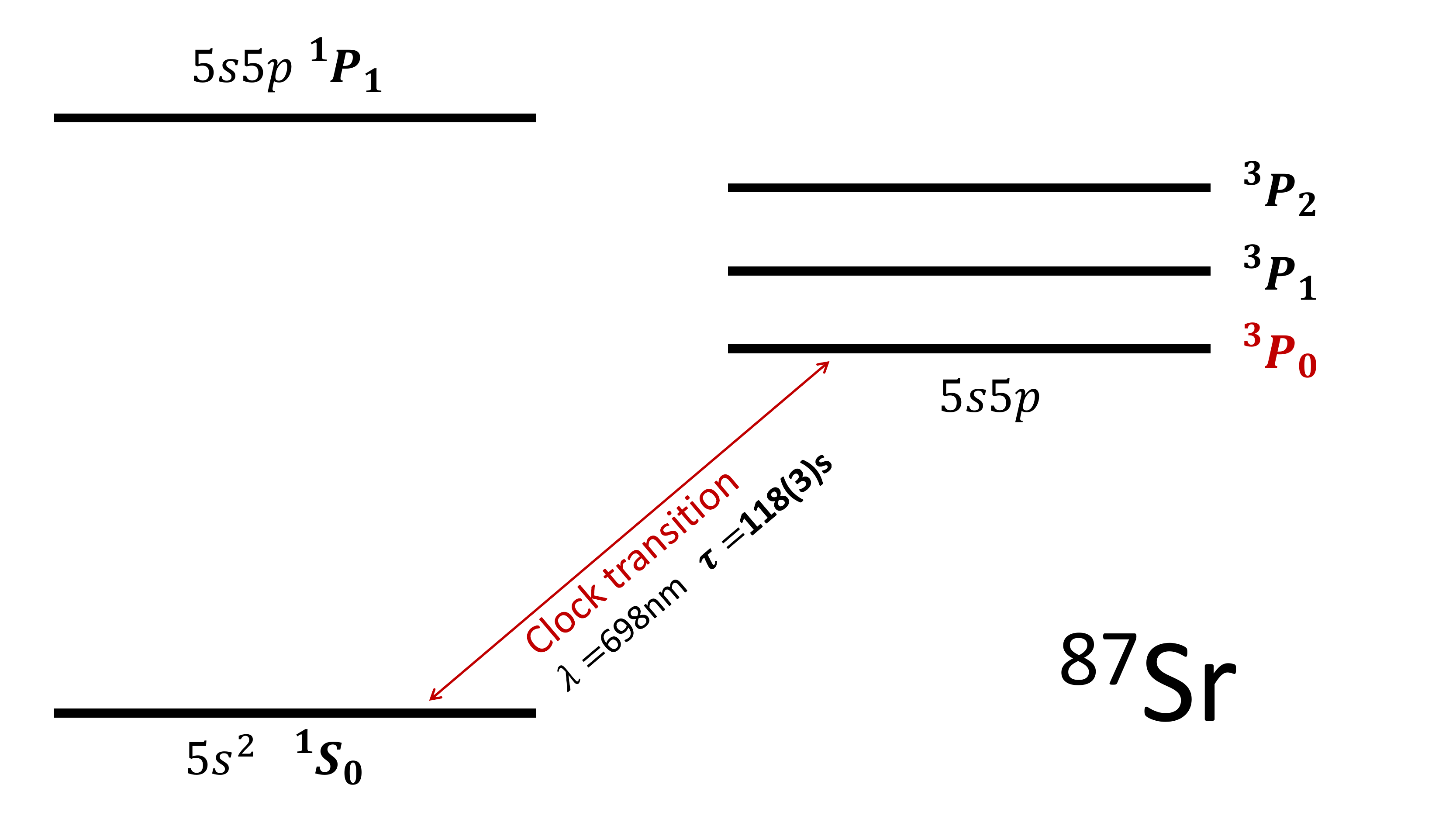}
    \caption{\it The clock transition in $^{87}$Sr that is favoured for attaining large LMTs in atom interferometers such as AION and MAGIS. The lifetime of the $^3P_0$ state of $\tau>100\,$s corresponds to an ultra-narrow transition with $\gamma_0/(2\pi)=1.35(3)\,$mHz~\cite{Muniz2021}.}
    \label{fig:StrontiumClockTransition}
\end{figure}

\paragraph{Spontaneous emission limit}
Spontaneous emission occurs when an excited atom spontaneously decays back to its ground state by emitting a photon and is a fundamental limitation in atom interferometry. The spontaneous emission process is stochastic, with the timing and direction of the photon emission being unpredictable. As a result, spontaneous emission introduces a random phase shift in the interferometer, which limits the precision of the measurement. For a given atomic transition, the amount of achievable LMT is ultimately limited by the rate of spontaneous emission, which is proportional to the inverse of the excited-state lifetime. 
As a consequence, many current proposals plan to use clock transitions, i.e., doubly-forbidden ultra-narrow transitions in atoms such as strontium (see Fig.~\ref{fig:StrontiumClockTransition}) or ytterbium that are very weak and hence provide mHz linewidths and long lifetimes of the excited states with lifetimes $>100\,$s. These transitions also form the basis of optical atomic clocks~\cite{Ludlow2015}. While the use of ultra-narrow clock lines avoids spontaneous emissions, it requires high laser powers to excite these weak lines at a suitable rate to achieve large LMT.

\paragraph{Squeezing}
As detailed above, when measuring the output of an interferometer, one characteristic limitation on the achievable precision arises from the quantum noise associated with the counting of uncorrelated probe particles. This shot noise or projection noise gives rise to the standard quantum limit (SQL) of phase resolution. It can in principle be reduced to the fundamental Heisenberg limit by entangling the probe particles, a process known as squeezing. 20dB of squeezing have recently been demonstrated in cold atoms with the help of an optical cavity, which would lead to a 100-fold reduction in phase noise~\cite{hosten2016measurement}.

\subsection{Signals of ultralight dark matter and gravitational waves in atom interferometers}

Here we describe briefly the possible signals of ultralight dark matter and gravitational waves in atom interferometers in gradiometer configurations.

\paragraph{Ultralight dark matter}
As discussed in more detail in Sec.\ \ref{DM}, scalar ultralight dark matter (ULDM) can be modelled as a temporally and spatially oscillating  classical field. 
If ULDM couples in a weak but finite non-gravitational linear way to Standard Model particles~\cite{Damour:2010rm,Damour:2010rp}, the presence of the dark matter field would manifest itself as a periodic modulation of the excitation energy and hence the transition frequency $\omega \to \omega + \Delta \omega$ between the atomic ground and excited states, $|g \rangle$ and $|e \rangle$, generating a phase shift in atom interferometers:
\begin{equation}
    \frac{1}{\sqrt{2}} |g \rangle + \frac{1}{\sqrt{2}} |e \rangle e^{-i\omega T} \quad \to \quad \frac{1}{\sqrt{2}} |g \rangle + \frac{1}{\sqrt{2}} |e \rangle e^{-i(\omega +\Delta \omega)T} \, .
\end{equation}
Spatial gradients of the dark matter field are expected to be too small to be detectable, but ULDM will nonetheless make its presence felt in an atom gradiometer, as the interferometer pulses interact with the different atomic clouds at different times separated by $\Delta t=c \Delta r$, giving rise to differential phase shifts between the interferometers.

\paragraph{Gravitational waves}
While ULDM is ultimately detected by its effect on the atomic transition frequencies, a passing gravitational wave (GW) is detected via the strain it creates in the space between the free-falling atoms. This strain changes the light propagation time, and therefore gives rise to an additional phase shift for the light pulses seen by the spatially-separated interferometers in the gradiometer:
\begin{equation}
    \frac{1}{\sqrt{2}} |g \rangle + \frac{1}{\sqrt{2}} |e \rangle e^{-i\omega T} \quad \to \quad \frac{1}{\sqrt{2}} |g \rangle + \frac{1}{\sqrt{2}} |e \rangle e^{-i\omega (T +\Delta T)} \, .
\end{equation}
In both cases, the resulting phase is directly proportional to the separation $\Delta r$ between the two interferometers and the number $n$ of LMT pulses applied. Hence, large-scale interferometers are required to reach the desired sensitivities.





\section{Landscape of Large-Scale Atom Interferometer Experiments}
\label{Landscape}

The experimental landscape of atom interferometry projects has expanded significantly in recent years, with several initiatives
underway to construct projects exploiting different cold atom technologies beyond the conventional laboratory environment. 

These include the construction of prototype projects at scales ${\cal O}(10 - 100)$~m, namely a 10-m fountain at Stanford and MAGIS-100 at FNAL in the US~\cite{MAGIS-100:2021etm}, AION-10 at Oxford with possible 100-m sites at Boulby in the UK and at CERN under investigation~\cite{AION}, MIGA in France~\cite{Canuel:2017rrp}, VLBAI at Hannover in Germany~\cite{schlippert2020matter} and a 10-m fountain and the ZAIGA project in China~\cite{Zhan:2019quq}.

These projects will demonstrate the feasibility of atom interferometry at macroscopic scales, paving the way for km-scale terrestrial experiments as the next steps. 
It is foreseen that by about 2035 one or more km-scale detectors will have entered operation. These km-scale experiments would not only be able to explore systematically for the first time the  mid-frequency band of gravitational waves ${\cal O}(10^{-2} - 10^{2})$~Hz, but would also serve as  technology readiness demonstrators for a space-based mission such as AEDGE~\cite{AEDGE} that would reach the ultimate sensitivity to explore the fundamental physics goals outlined in this article.
     
In the following Sections~\ref{Terrestrial} and \ref{Space} we outline further details of the ongoing and proposed terrestrial projects and the space-based mission concept, respectively, with a view to highlighting the enormous science potential of large-scale atom interferometry projects.   
\subsection{Terrestrial}
\label{Terrestrial}

The expanded landscape of terrestrial atom interferometer experiments includes projects ranging from ultra-sensitive laboratory setups to portable devices and commercially available gravimeters. These developments have been made possible by advancements in technology, including improved laser cooling and trapping techniques, better control over atom interferometer systems, and more precise readout and feedback mechanisms.
We now review the larger terrestrial atom interferometer experiments ranging in size
from ${\cal O}(10)$~m to ${\cal O}(1)$~km that are in operation, in preparation,
planned or proposed in different geographical locations around the world.

\begin{figure}
    \centering
    \includegraphics[width=10cm]{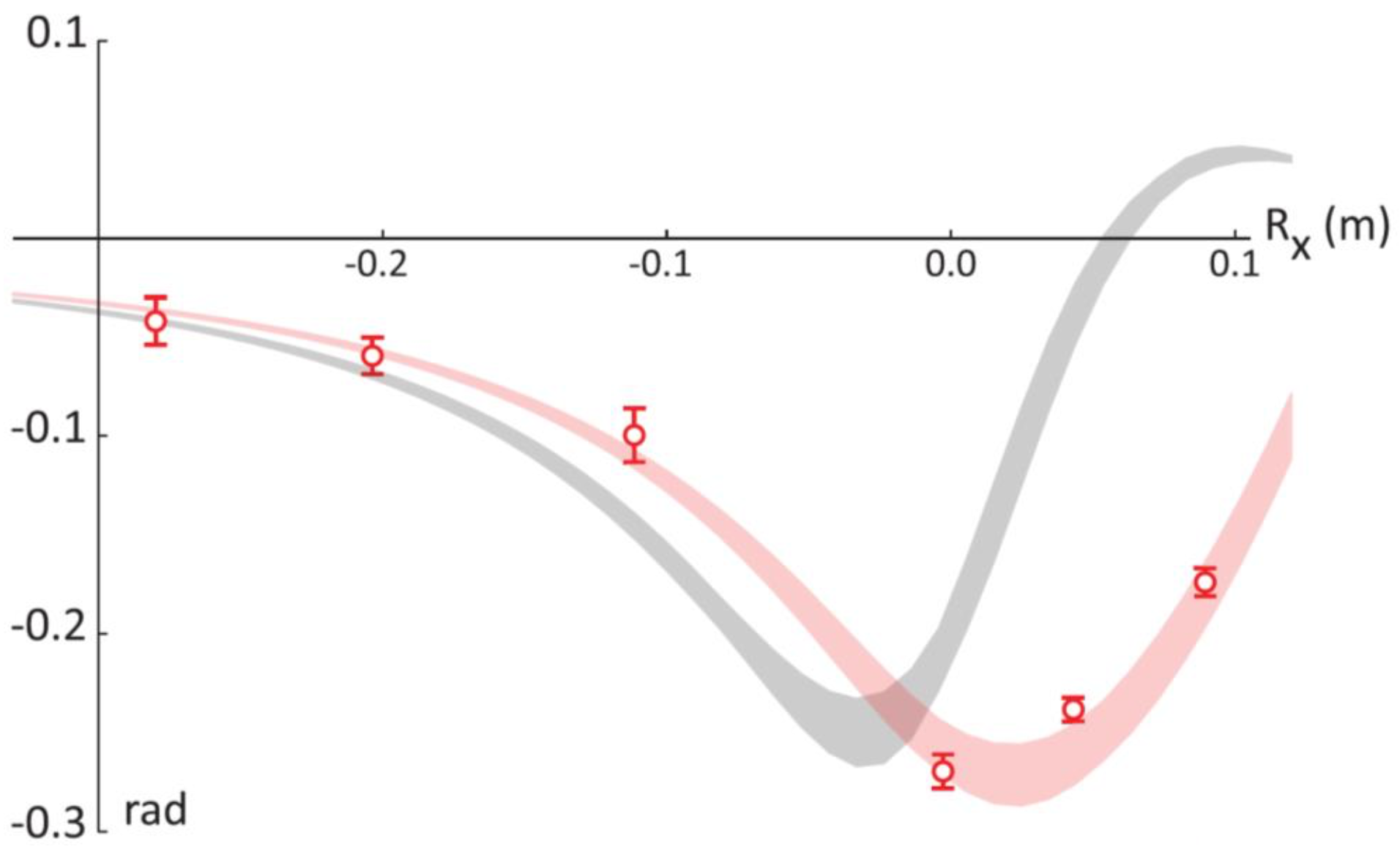}
\caption{\it Phase shift induced by a tungsten source mass adjacent to an atomic gradiometer using $^{87}$Rb clouds (red points) compared with theoretical predictions based on a semiclassical quantum-mechanical calculation (pink band). 
The grey band is based on a calculation neglecting the Gravitational Ahoronov-Bohm effect: see~\cite{Overstreet:2021hea} for details.}
\label{fig:GravitationalAhoronovBohm}
\end{figure}

In the US, a pioneering facility has been operating for some years at Stanford University~\cite{Asenbaum:2016djh}.
It uses a 10-m vertical vacuum chamber into which clouds of rubidium are launched. One of
the experiments performed in this facility has been to test the Einstein Equivalence Principle
(EEP) by comparing the motion in free fall of clouds of $^{85}$Rb and $^{87}$Rb. The EEP
was verified at the $10^{-12}$ level~\cite{Asenbaum:2020era}, the most precise check in a terrestrial experiment.
Another experiment used pairs of $^{87}$Rb clouds launched 
simultaneously to different heights
to verify a gravitational analogue of the Aharonov-Bohm effect~\cite{Overstreet:2021hea}, namely a phase shift
\begin{equation}
    \Delta \phi \; = \; m \int [V(x_1,t) - V(x_2,t)] dt
\end{equation}
induced by the gravitational field of a tungsten source mass placed close to one of the clouds, as depicted in
Fig.~\ref{fig:GravitationalAhoronovBohm}.

\begin{figure}
    \centering
    \includegraphics[width=14cm]{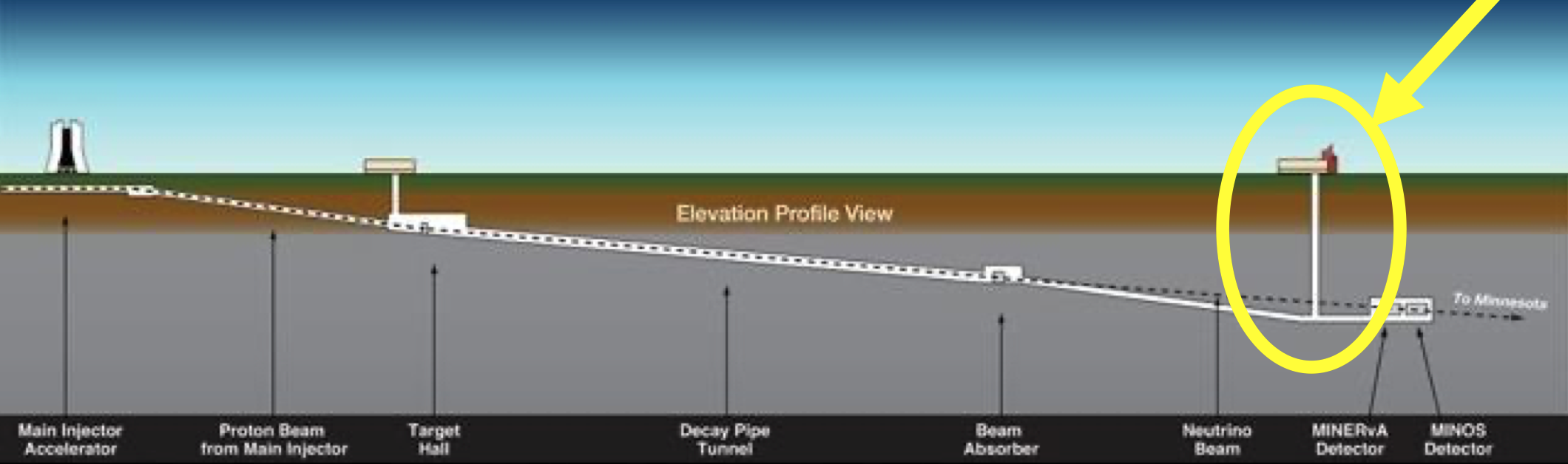}
    \caption{\it Location of the MAGIS-100 experiment (highlighted) in a vertical access shaft to the long-baseline neutrino beam at Fermilab~~\cite{MAGIS-100:2021etm}.}
    \label{fig:MAGIS100}
\end{figure}
Another 10-m fountain using strontium is currently under construction in another Stanford laboratory,
and a next-stage US experiment, MAGIS-100, is currently being prepared for installation in a 100-m
vertical shaft at Fermilab~\cite{MAGIS-100:2021etm}, see Fig.~\ref{fig:MAGIS100}. It plans to use clouds of strontium to probe the possible couplings
of ultralight dark matter (ULDM) to Standard Model particles, and to act as a pathfinder for the search
for GWs. A follow-on strontium experiment could be located in a 2-km vertical shaft at
the Sanford Underground Research Facility (SURF) in South Dakota~\cite{heise2022sanford}. In addition to extending the
search for ULDM interactions, it would pioneer the search for GWs in the range of frequencies
intermediate between LIGO/Virgo and LISA.

\begin{figure}
    \centering
    \includegraphics[width=7cm]{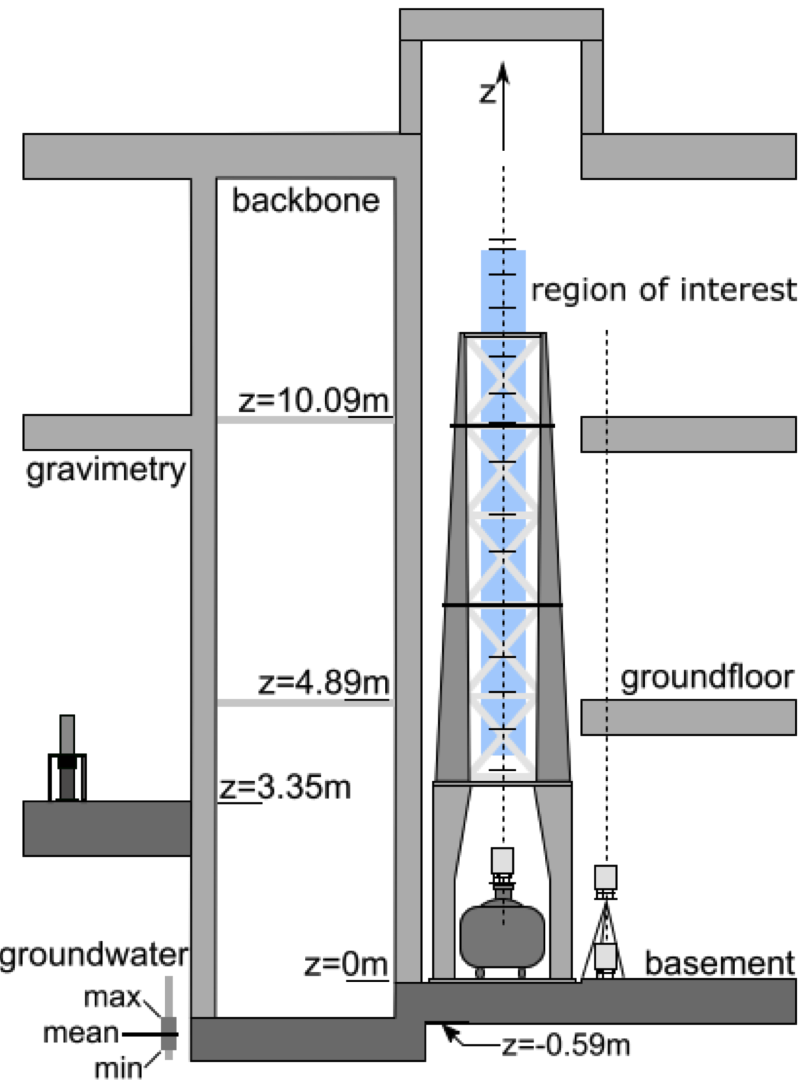}
    \hspace{0.5cm}\vspace{0.25cm}
    \includegraphics[width=7cm]{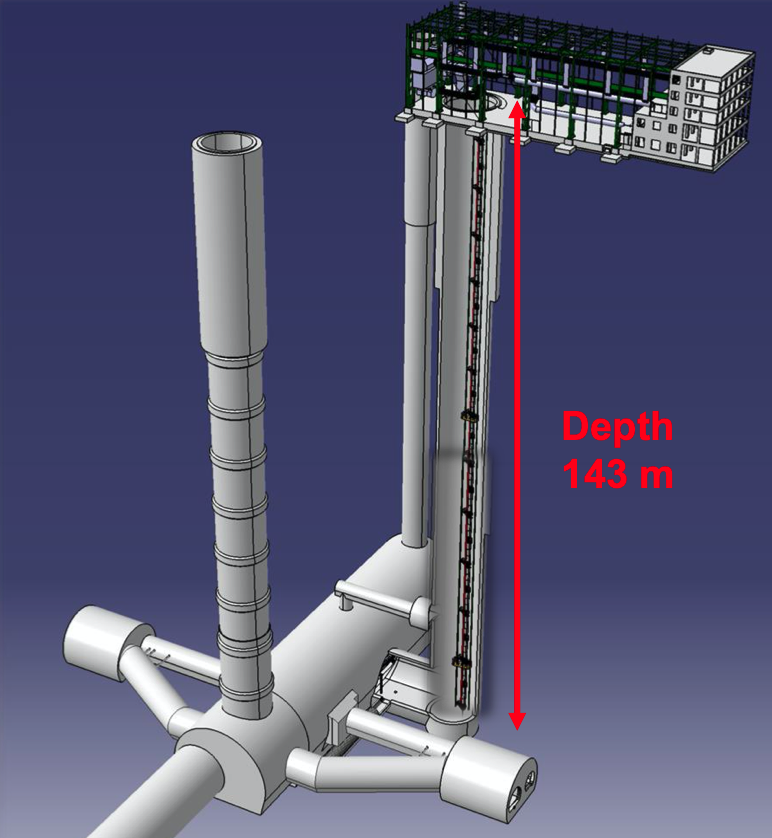}\\
    \vspace{0.5cm}
    \includegraphics[width=7cm]{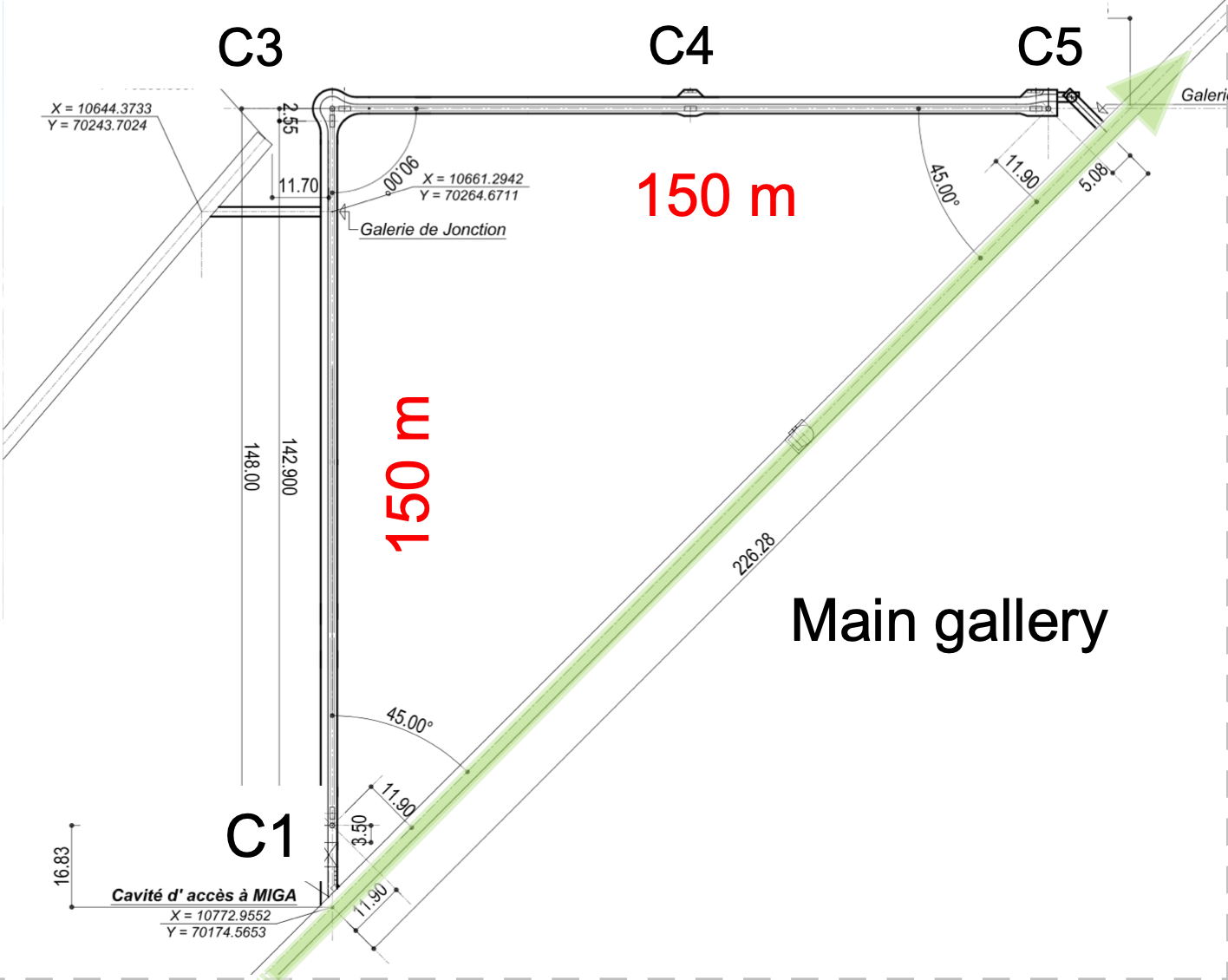}
    \hspace{0.5cm}
    \includegraphics[width=7cm]{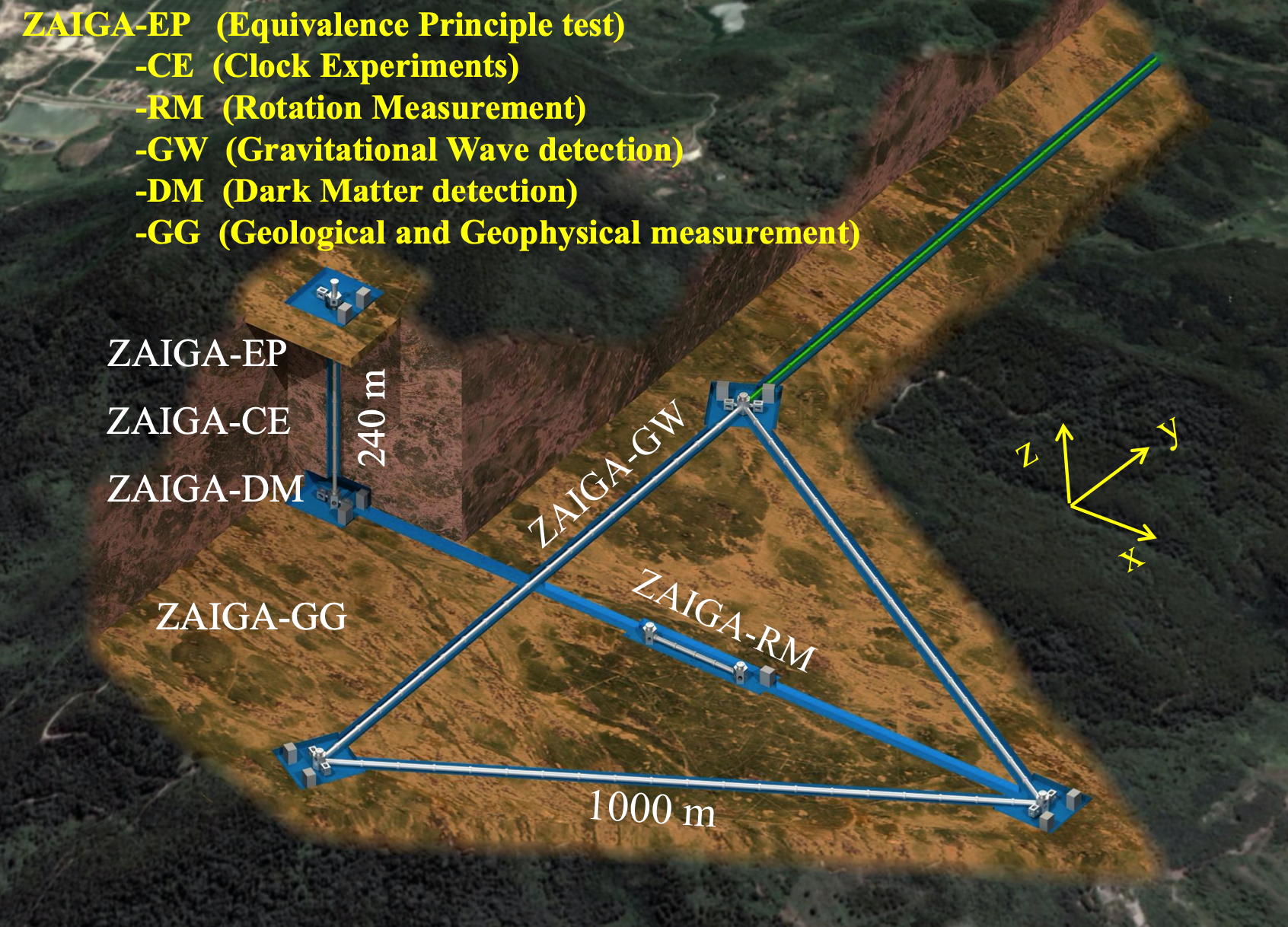}
    \caption{\it Upper left: Design of the 10-m VLBAI experiment that is currently under construction
in Hannover~\cite{schlippert2020matter}.
Upper right: Possible location of a vertical 100-m atom interferometer in an LHC access shaft at CERN~\cite{Arduini:2023wce}.
Lower left: Layout of the MIGA experiment in the Laboratoire souterrain {\` a} bas bruit (LSBB) in Rustrel, France~\cite{Canuel:2017rrp}.
Lower right: Layout of the ZAIGA laboratory near Wuhan, China for a range of experiments using atom interferometry~\cite{Zhan:2019quq}.}
    \label{fig:Hannover}
\end{figure}
In Germany, the Very Long Baseline Atom Interferometry (VLBAI) test stand experiment is under construction
in Hannover~\cite{schlippert2020matter}, see the upper left panel of Fig.~\ref{fig:Hannover}. It will consist of a 10-m vertical atom interferometer operated as a gravimeter using 
mixtures of ultracold ytterbium and rubidium atoms. In addition to high-precision matter-wave gravimetry
and gradiometry, it will make a quantum test of the EEP, investigate decoherence and dephasing in 
atom interferometers and perform research on quantum clocks.

\begin{figure}
    \centering
    \includegraphics[width=2.5cm]{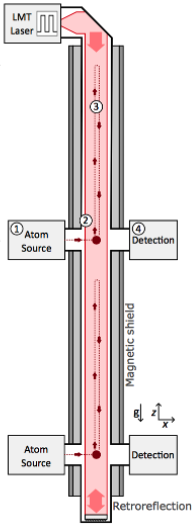}
    \hspace{1cm}
    \includegraphics[width=7cm]{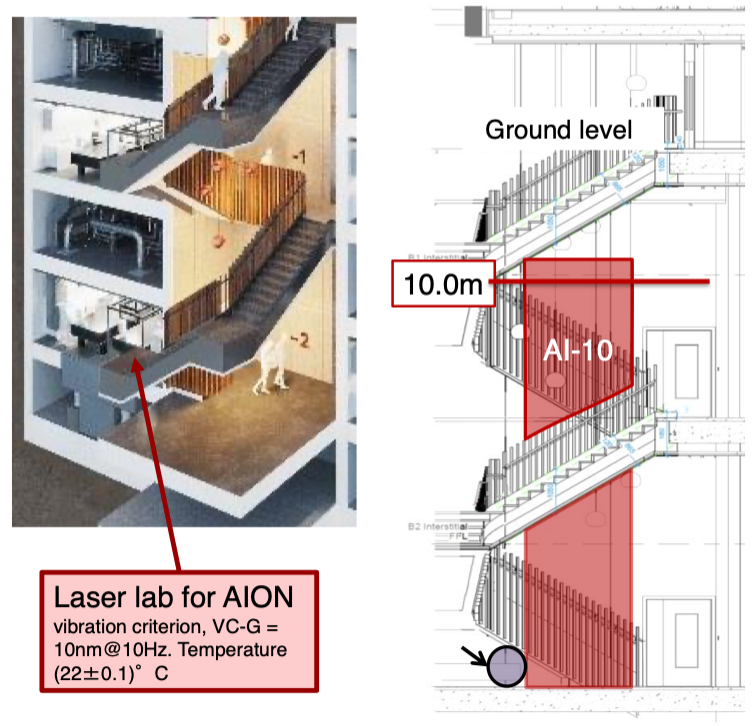}
    \caption{\it Left panel: Schematic drawing of an atom gradiometer using two atom sources (1) that launch clouds of strontium atoms in a common vertical tube (2) and addresses them with a common interferometer laser (3), before detecting their interference patterns (4). Right panels: Layout of the AION-10 atom interferometer in the basement of the Oxford Physics Department.}
    \label{fig:Oxford}
\end{figure}

In the UK, the Atom Interferometry Observatory and Network (AION) project~\cite{AION} envisages a staged series of
atom interferometers using fountains of strontium clouds in vertical vacuum chambers, as illustrated schematically in the left panel of Fig.~\ref{fig:Oxford}. The first stage is to
build a 10-m device~\footnote{See~\cite{AION:2023fpx} for a summary of the current technical status of the AION project.}  in the basement of the Oxford University Physics Department, also illustrated in Fig.~\ref{fig:Oxford}, which would make
initial probes of possible ULDM couplings to Standard Model particles.
This would be followed by a
100-m detector that could be located either in the UKRI Boulby Underground Laboratory or at CERN - see the upper right 
panel of Fig.~\ref{fig:Hannover}.
AION-100 would probe further such ULDM couplings and make initial searches for GWs.
These AION detectors would be operated in partnership with the MAGIS-100 detector, with which they shares many
common technologies. The final terrestrial step for AION would be a km-scale detector, which could be
installed near the 100-m detector in the UKRI Boulby Underground Laboratory. The fourth step
in the AION programme would be the AEDGE space-borne detector discussed in Section~\ref{Space}.

In France, the Matter-wave laser Interferometric Gravitation Antenna (MIGA) is to be installed in the Laboratoire souterrain {\` a} bas bruit (LSBB) in Rustrel at a depth of 300~m~\cite{Canuel:2017rrp}, see the lower left panel of Fig.~\ref{fig:Hannover}. It will have two orthogonal horizontal 150-m arms, each containing a pair of parallel laser beams that drive three rubidium interferometers, to measure the components of the gravitational field. The differential measurements of the gravitational field will enable the local gravity gradient to be extracted. MIGA is intended to pave the way towards ELGAR~\cite{ELGAR}, a proposed European research infrastructure with two horizontal 16-km arms that could study gravity with unprecedented precision.

In China, the Zhaoshan long-baseline Atom Interferometer Gravitation Antenna (ZAIGA) is a large-scale interferometer facility that is currently under construction at an average depth of 200~m under a mountain near Wuhan~\cite{Zhan:2019quq}. ZAIGA will combine long-baseline atom interferometers, high-precision atom clocks, and large-scale gyros, as seen in the lower right panel of Fig.~\ref{fig:Hannover}. It will combine a horizontal equilateral triangle of tunnels with two rubidium interferometers separated by 1~km in each arm with a 300-m vertical shaft and another horizontal 1-km tunnel housing optical clocks linked by locked lasers. The ZAIGA facility will be used for gravitational wave detection, a test of the EEP, and measurements of the gravitational red-shift, rotation and the gravito-magnetic effect.

The goal of the international community interested in long-baseline terrestrial atom interferometer experiments is to have at least one km-scale detector in operation by 2035, which would maximise the possible terrestrial exploration of the mid-frequency band of gravitational waves as well as probe possible ultralight dark matter. It would also demonstrate the readiness of key technologies ahead of a space-based atom interferometry mission such as AEDGE (see Section~\ref{Space}).
With this goal in mind, a workshop was organized at CERN to develop a roadmap for the design and technology choices for one or several km-scale detectors to be ready for operation in the mid-2030s~\cite{TVLBAI}. This workshop brought together the cold atom, astrophysics, cosmology, and fundamental physics communities and built upon the previous Community Workshop on Cold Atoms in Space held in September 2021, which established a corresponding roadmap for cold atoms in space~\cite{Alonso:2022oot}. The workshop laid the basis for a roadmap for terrestrial experiments outlining technological milestones and refining the interim and long-term scientific goals that would guide the development of km-scale atom interferometers.

Participants in the workshop agreed that the global community interested in such experiments would work together towards the establishment of an informal proto-collaboration that could develop the science case for such facilities and provide a forum for exchanging ides how to develop the necessary technological advances,
as well as coordinate implementation of the roadmap for their realisation.

\subsection{Space}
\label{Space}

Atomic Experiment for Dark Matter and Gravity Exploration (AEDGE) is a concept for a space mission deploying two strontium atom interferometers in a pair of satellites in medium earth orbit with a nominal separation of $4 \times 10^4$~km~\cite{AEDGE}. It was proposed to the European Space Agency (ESA) in response to its Voyage 2050 call for scientific mission concepts to follow the launch of the LISA GW experiment. AEDGE would measure GWs in the mid-frequency band intermediate between LISA and terrestrial laser interferometers such as LIGO, Virgo, KAGRA, ET and CE. This range of frequencies could be expanded if it is found to be feasible to use atom clouds outside the spacecraft, a configuration called AEDGE+~\cite{Badurina:2021rgt}.
The AEDGE concept was received favourably by an ESA senior advisory committee, but the need for a preparatory programme of technology development and space qualification was emphasised. 

\begin{figure}[htb]
\centering 
\includegraphics[width=12cm]{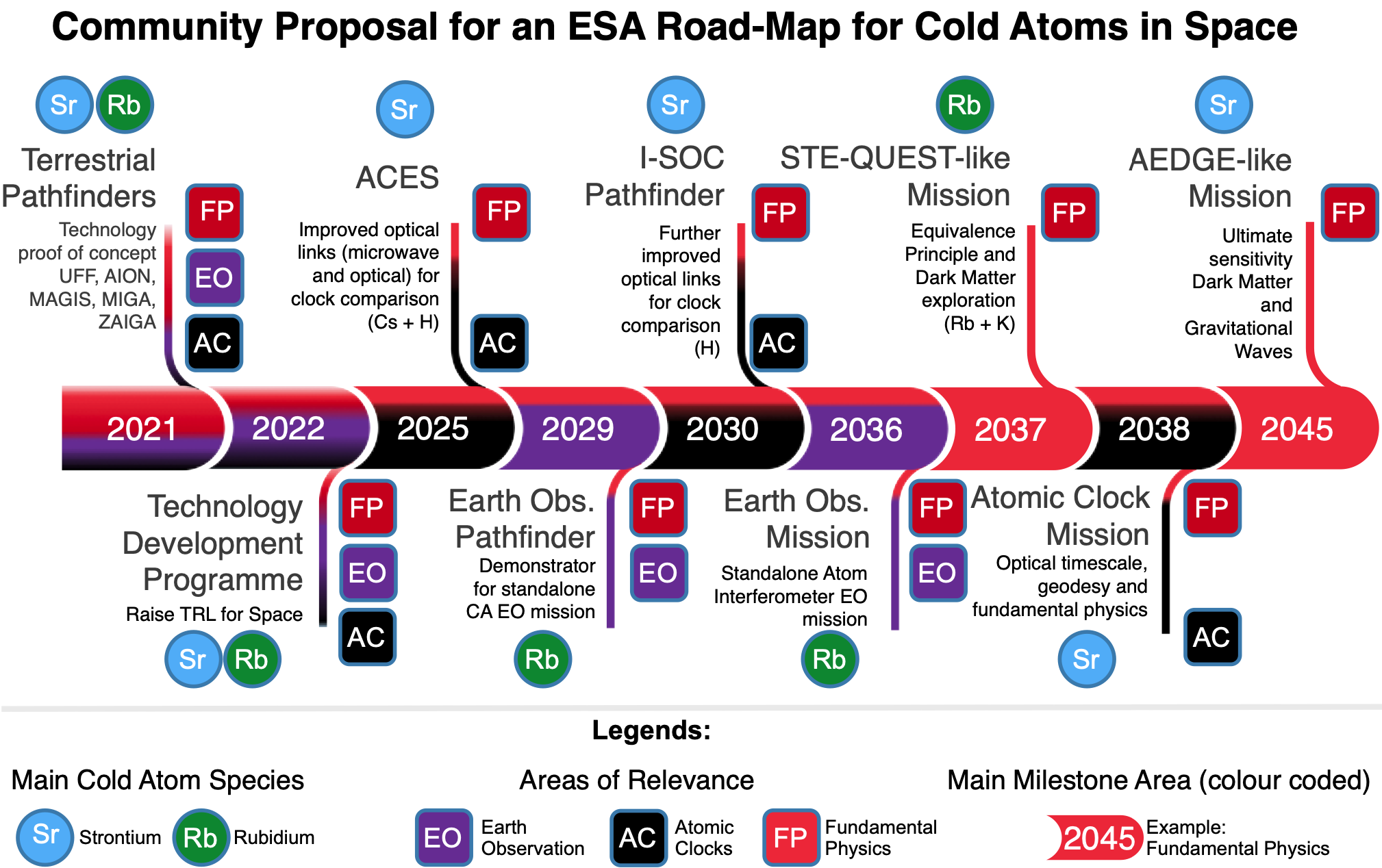}
\caption{\it A community roadmap for cold atoms in space, taken from~\cite{Alonso:2022oot}. The main milestones of the roadmap are colour coded according to relevance for the three main areas of Earth Observation (purple), Atomic Clocks (black) and Fundamental Physics (red). The main cold atom species targeted by the milestones are indicated with circles coloured blue for strontium and green for rubidium. The year indicated for each milestone represents its starting date or the launch date of a mission. Further details about each of the milestones and their interdependences are given in~\cite{Alonso:2022oot}.}
\label{fig:CASpaceRoadmap}
\end{figure}

The Community Workshop on Cold Atoms in Space in 2021 was organised in response to provide an opportunity for experts from various communities to discuss together the development of a cold atom quantum technology programme coordinated at the European level. The workshop outlined a roadmap for future developments in this field with the aims of enhancing the strength and coherence of a community embracing cold atom technology experts as well as prospective users. This was presented in a community report on the prospects for cold atoms in space~\cite{Alonso:2022oot}, see Fig.~\ref{fig:CASpaceRoadmap}, which highlighted the synergies and technological commonalities between space projects for fundamental science and next-generation missions for earth observation, geodesy and time-keeping. 

Current  space-borne R\&D work towards large-scale mission concepts includes pathfinder missions deploying cold atom experiments, such as CACES~\cite{Liu:2018kn}, MAIUS~\cite{MAIUS} and CAL~\cite{CAL}. Prototypes for the key underlying optical technologies such as FOKUS~\cite{Lezius}, KALEXUS~\cite{Dinkelaker} and JOKARUS~\cite{PhysRevApplied.11.054068} have already demonstrated reliable operation, and much more space experience will be gained in the coming years. In parallel, several other cold atom projects are aiming to demonstrate the general space-readiness of cold-atom technology including the scaling of the basic parameters that are required for large-scale space-borne mission concepts, e.g., the medium-scale STE-QUEST mission concept~\cite{Tino2019,Ahlers:2022jdt}.

These and other technology demonstrators and pathfinder missions will serve as milestones that help to test and validate cold-atom technology, as well as identify any potential issues that need to be addressed before full-scale missions such as AEDGE can be launched. The roadmap document~\cite{Alonso:2022oot} provides detailed requirements for technology development and space qualification, and reviews pathfinder missions in detail. This information provided serves to guide researchers and developers who are working on cold atom technologies and will help to ensure that they meet the necessary standards for use in space.
The proposed roadmap is in synergy with European Union (EU) programmes and recommendations from the ESA Voyage 2050 Senior Science Committee. This alignment will help to ensure that efforts are coordinated across different organizations and that resources are used efficiently.

\section{Fundamental Physics with Large-Scale Atom Interferometers}
\label{Applications}

\subsection{Dark Matter}
\label{DM}

Despite the overwhelming astrophysical and cosmological evidence for the gravitational effects of cold (non-relativistic) dark matter, its composition is mysterious. Searches for WIMPs continue at the LHC and via direct and indirect astrophysical signals, but without success so far, see, e.g., \cite{ATLAS:2022ihe,CMS:2023ktc,LZ:2022ufs}. This has led to increasing interest in ultra-light dark matter (ULDM) in the form of coherent waves of scalar bosons such as dilatons, moduli or the relaxion, pseudoscalar axion-like particles, vector or tensor bosons. Atom interferometers have unique capabilities to detect ULDM  candidates that couple to SM particles, and proposals have been made for probing scalar, pseudoscalar and vector fields of dark matter~\cite{Graham:2015ifn, Arvanitaki:2016fyj}.


A coupling between scalar ultra-light dark matter (ULDM) and electrons, photons, or quarks can produce a time-varying modification of the effective mass of electrons, fine-structure constant of electromagnetism, or mass of quarks, respectively~\cite{Stadnik:2014tta}. The time variations in these parameters would lead to small modulations in the energy levels of atoms, including the ``clock" transition in strontium, see Fig.~\ref{fig:StrontiumClockTransition}, which is utilised in experiments such as AION, MAGIS, and AEDGE.

The simplest possibility is that the scalar ULDM field $\phi(\mathbf{x},t)$ couples to SM fields~\cite{Damour:2010rm,Damour:2010rp} through {\it linear} interactions of the form
\begin{equation}
\mathcal{L}^{\rm{lin}}_{\rm{int}}  \supset - \phi(\mathbf{x},t) \cdot \sqrt{4 \pi G_{\rm{N}}} \cdot \left[ d_{m_e} m_e \bar{e}e - \frac{1}{4} d_e F_{\mu \nu} F^{\mu \nu} + d_{m_q} m_q \bar{q}q\right]  + b\, \phi(\mathbf{x},t) |H|^2    \;,
\label{linear}
\end{equation}
where $G_{\rm{N}}$ is Newton's constant, $m_{e,q}$ are the electron and quark masses, $d_{m_{e,q}}$, $d_e$ and $b$ are the couplings of the scalar ULDM to normal matter, and we have used units where $\hbar=c=1$.
The large ULDM occupation number implies that the scalar ULDM field behaves as a non-relativistic oscillating field approximated by
\begin{equation}
\phi(\mathbf{x},t)=\frac{\sqrt{2 \rho_{\rm{DM}}}}{m_{\phi}} \cos[m_{\phi}(t - \mathbf{v}_{\phi}\cdot \mathbf{x})+\cdots]\;,
\end{equation}
where $m_{\phi}$ is the scalar DM mass, $\rho_{\mathrm{DM}}$ is the 
local cold DM density, whose average value we take to be $\sim 0.3~\mathrm{GeV}/\mathrm{cm}^3$, and $\mathbf{v}_{\phi}$ is the DM velocity whose magnitude $|\mathbf{v}_{\phi}|$ and dispersion $v_{\rm{vir}}$ have values that are characteristically  $\sim 10^{-3}c$~\cite{Bertone:2016nfn}. The ellipses in the argument of the cosine include 
an unknown random phase $\theta$ that encodes the coherence properties of the ULDM field. Its coherence length is~\cite{Derevianko:2016vpm}
\begin{equation}
    \lambda_c = \frac{\hbar}{m_{\phi} v_{\rm{vir}}} \approx 2.0 \times 10^3 \,  \left( \frac{10^{-10}~\!{\rm eV} }{m_{\phi}}  \right) \, {\rm km} \, , 
\end{equation}
and its coherence time is
\begin{equation}
 \tau_c = \frac{\hbar}{m_{\phi} v_{\rm{vir}}^2} \approx 6.6 \,  \left(  \frac{ 10^{-10}~\!{\rm eV}  }{m_{\phi}} \right)\, {\rm s} \, .
\end{equation}
The magnitude of the oscillation in transition energy induced by ULDM depends on several factors, including the ULDM mass, the energy density of ULDM at the detector's location, and the strength of the linear couplings between ULDM and SM fields. The frequency of the transition-energy oscillation is determined by the ULDM mass, and the highest signal from ULDM-induced oscillation is observed when this frequency falls within the mid-frequency range commonly used by atom interferometer experiments.

\begin{figure}
\centering 
\includegraphics[width=7cm]{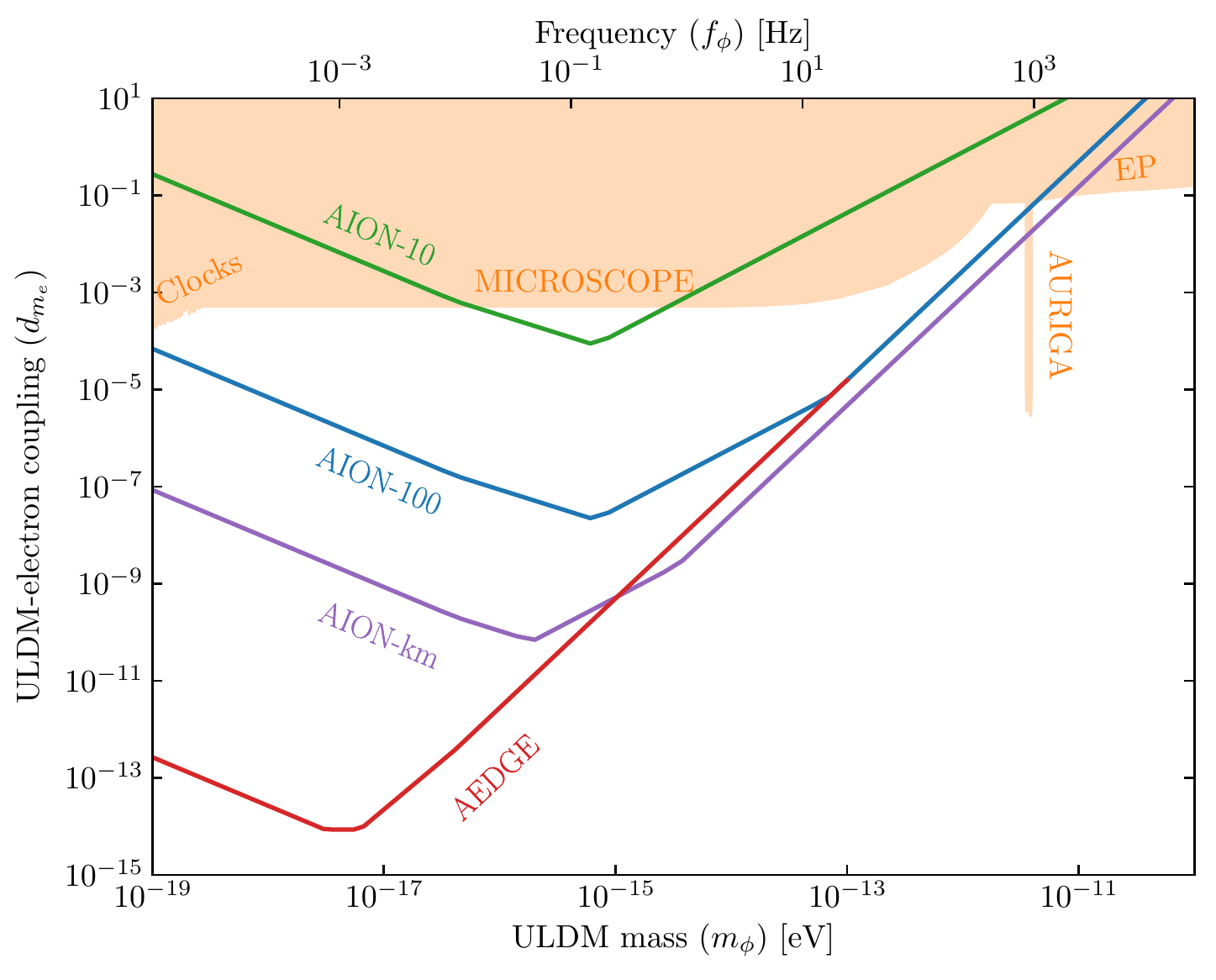}
\includegraphics[width=7cm]{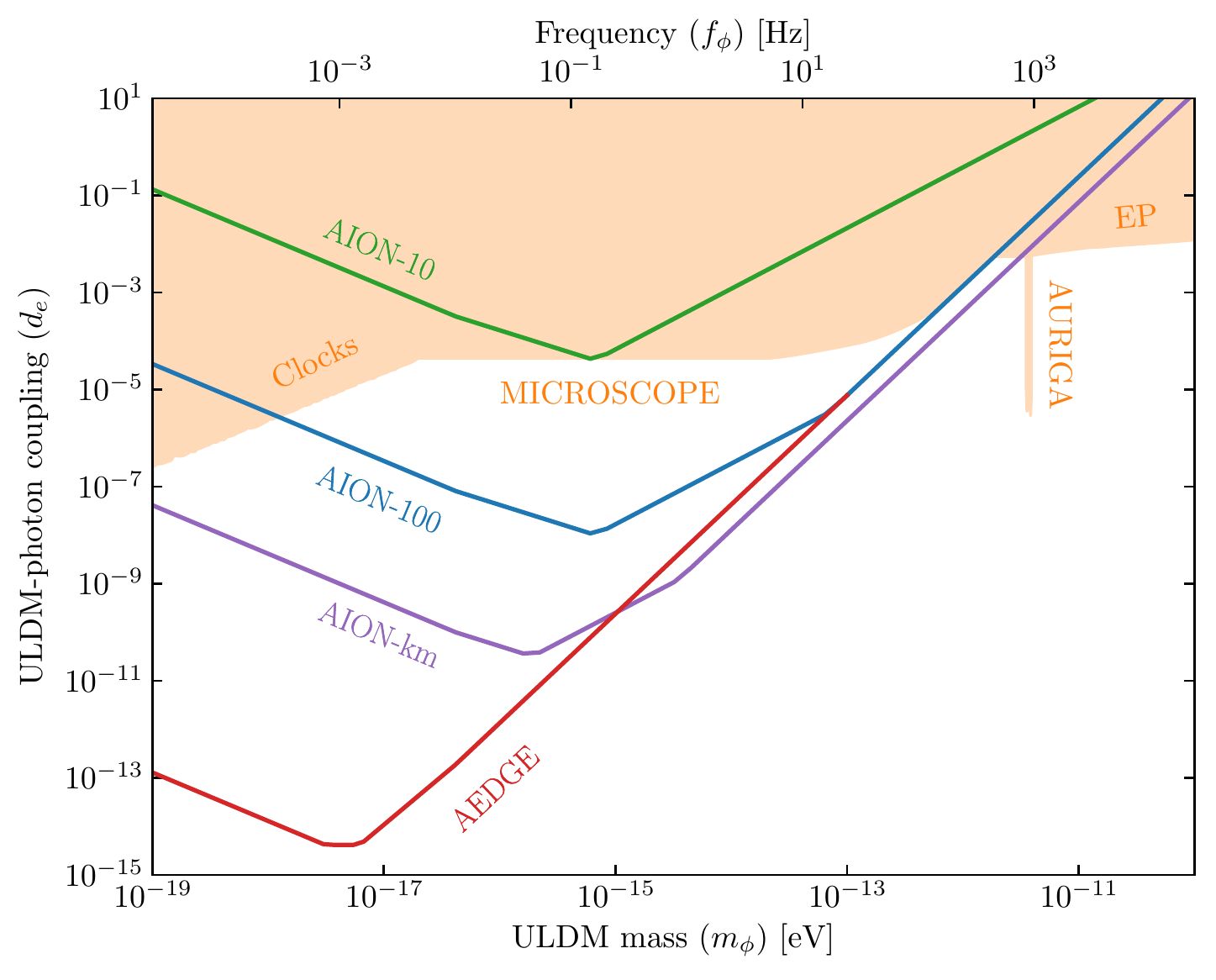}
\vspace{-3mm}
\caption{\it Sensitivity projections for linear ULDM couplings to electrons (left panel) and photons (right panel) with parameters given in~\cite{AION}.
The green (blue) (purple) and red lines are for AION-10 (-100) (-km) and AEDGE, respectively. The shaded orange region is excluded by the existing constraints from searches for violations of the equivalence principle by the MICROSCOPE experiment~\cite{MICROSCOPE:2022doy} and with torsion balances (labelled EP)~\cite{Wagner:2012ui}, atomic clocks~\cite{Hees:2016gop}, and the AURIGA experiment~\cite{Branca:2016rez}.}
\label{fig:AIONULDM}
\end{figure}

The panels in Fig.~\ref{fig:AIONULDM} display sensitivity projections for scalar ULDM that is linearly coupled to electrons (photons) with SNR = 1. These projections were computed assuming the experimental parameters given in Ref.\cite{AION}. The AION and AEDGE sensitivity curves are overlaid on existing constraints, which are indicated by orange shading. The atom interferometer sensitivity typically demonstrates oscillations with respect to the ULDM mass.
For clarity, only the envelope of the oscillations is plotted in the AION-100, AION-km, and AEDGE projections. Additionally, the AEDGE sensitivity curve is shown only until the point at which it approaches the AION-km line, although its sensitivity also extends to higher frequencies, where it would overlap with the AION-km line.


Figure~\ref{fig:AIONULDM} demonstrates the potential of atom interferometers for exploring uncharted regions of parameter space in the ULDM couplings to SM fields for scalar ULDM masses ranging from $10^{-18}$~eV to $10^{-12}$~eV. AION-10 aims to achieve, or even exceed, existing constraints, while AION-100 and AION-km are expected to expand significantly the range of detectable couplings to lower values. It should be noted that these projections assume atom shot-noise as the limiting factor for phase-noise. However, it is expected that below approximately 0.1~Hz Gravity Gradient Noise (GGN) will become the dominant noise source. The magnitude of the GGN is site-dependent and may be mitigated to some extent by the experimental design: see the discussion in~\cite{Badurina:2022ngn}. In contrast, space-borne experiments are not subject to GGN, and the AEDGE projections are not subject to such uncertainties at lower frequencies, corresponding to lower ULDM masses.

\subsection{Gravitational Waves (GWs)}
\label{GWs}

The discovery of GWs by the LIGO and Virgo experiments not only confirmed a century-old prediction
by Einstein, but also opened a new window on the Universe through which hitherto invisible objects
could be detected. {\it A priori}, the GW spectrum could be as rich as the electromagnetic
spectrum, and its exploration will require a combination of experimental techniques that are
optimised for different frequency ranges. The terrestrial LIGO, Virgo and KAGRA laser interferometers
are optimised for GWs with frequencies that are ${\cal O}(100)$~Hz, the planned space-borne LISA
laser interferometer is optimised for frequencies that are ${\cal O}(10^{-2} - 10^{-4})$~Hz, and
Pulsar Timing Arrays (PTAs) and SKA~\cite{Janssen:2014dka} are in principle sensitive to GWs with frequencies that are ${\cal O}(10^{-9})$~Hz: see~\cite{Ellis:2023owy} and references therein.
There are gaps in the spectrum between the ranges where these detectors are sensitive, and atom interferometers
target primarily the ${\cal O}(10^{-2} - 100)$~Hz frequency band that is intermediate between those
targeted by LIGO/Virgo/KAGRA and LISA.

LIGO and Virgo have discovered GWs emitted by the mergers of black holes (BHs) weighing up to a few dozen solar masses,
as well as mergers of such BHs with neutron stars (NSs), and have also observed a couple of NS-NS mergers~\cite{KAGRA:2021duu}. More recently, the NANOGrav Collaboration has reported evidence for GWs in the nHz range~\cite{NANOGrav:2023gor} that may come from binary systems containing supermassive BHs (SMBHs)~\cite{Ellis:2023dgf} or from cosmic string networks~\cite{Ellis:2023tsl}.
So far, these observations are consistent with predictions derived from General Relativity, and the
measurements of NS-NS mergers begin to constrain models of the NS equation of state. On the other hand,
the observations of BH-BH mergers pose a number of astrophysical issues. What are the origins of the
detected BHs? Could they be primordial? Are there BHs or other compact objects with masses between 2 and 5
solar masses? Is there a BH mass gap around 100 solar masses, as predicted by models of stellar evolution?
Are there intermediate-mass BHs (IMBHs)~\cite{Greene:2019vlv} with masses intermediate between those detected by LIGO/Virgo/KAGRA and the SMBHs in the cores of galaxies? How were the SMBHs assembled? Answering the latter questions will require
measurements in the ${\cal O}(10^{-2} - 100)$~Hz frequency band targeted by large-scale atom interferometers.

\begin{figure}
\centering 
\includegraphics[width=9cm]{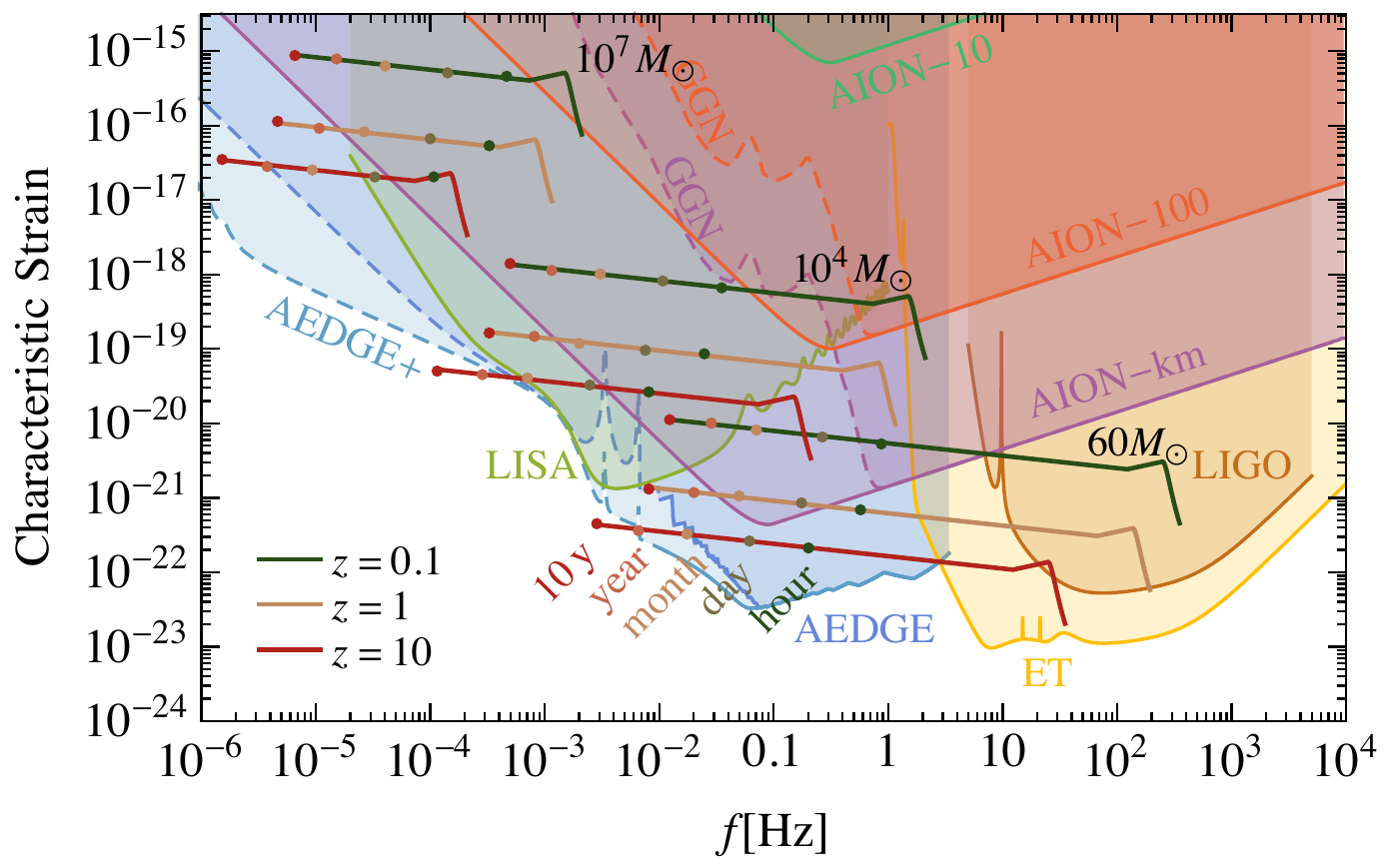}
\includegraphics[width=5.5cm]{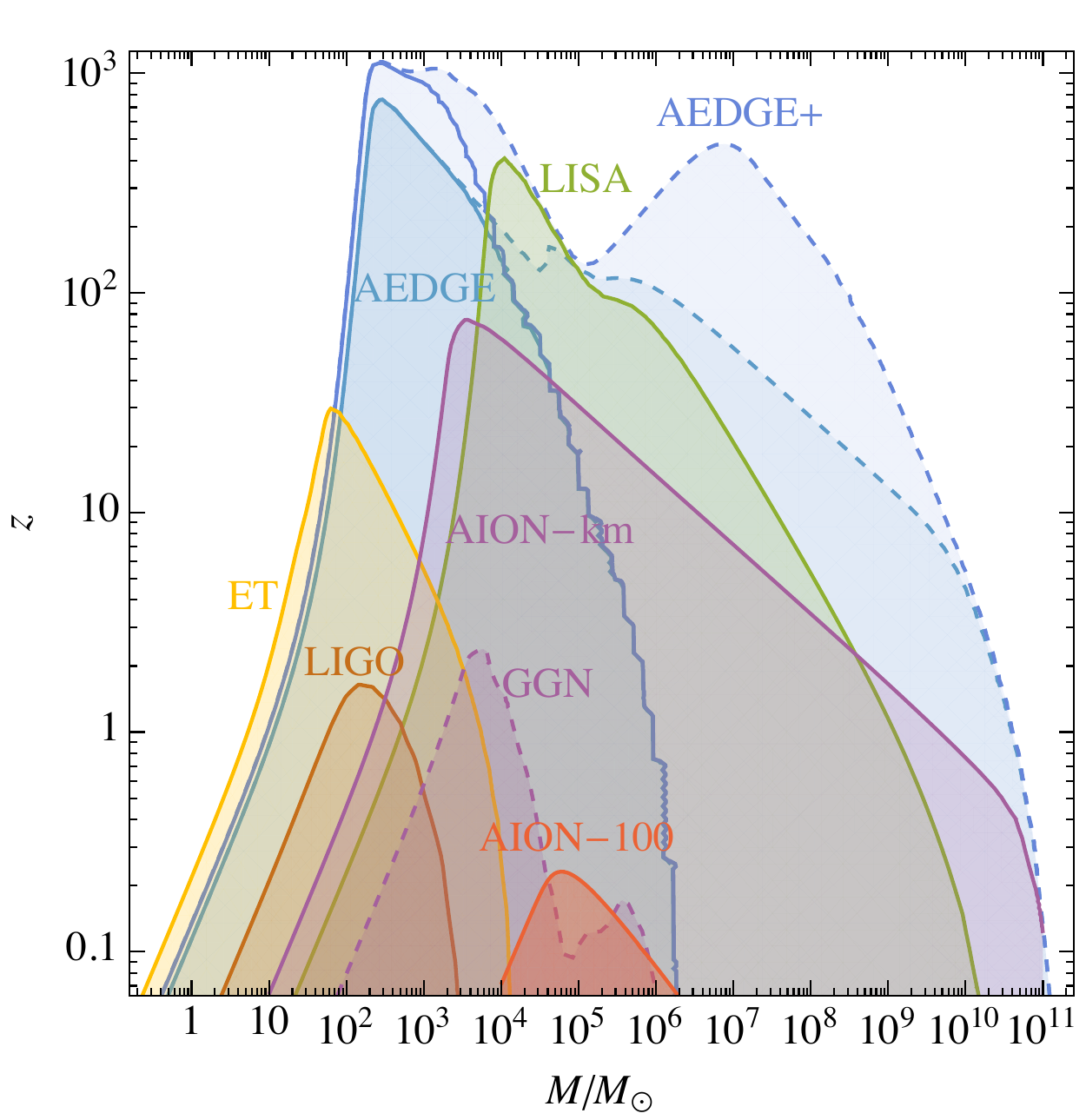}
\caption{\it Left panel: Strain sensitivities of AION-10, -100 and -km, AEDGE and AEDGE+, compared
with those of LIGO, LISA and ET and the signals expected from mergers of equal-mass
binaries whose masses are $60, 10^4$ and $10^7$ solar masses.  The assumed redshifts are
$z = 0.1, 1$ and $10$, as indicated. Also indicated are the remaining times during inspiral before the
final mergers. Right panel: Signal-to-noise ratio (SNR) $= 8$ sensitivities of LIGO, ET, LISA, AION
and AEDGE to equal-mass black hole binaries as functions of the total binary mass and the redshift $z$.
From~\cite{Badurina:2021rgt}.}
\label{fig:GWs}
\end{figure}

The left panel of Fig.~\ref{fig:GWs} shows as solid lines the possible GW sensitivities of proposed terrestrial atom interferometers 
(AION-10, -100 and -km)~\cite{AION} and the proposed space-borne atom interferometer (AEDGE)~\cite{AEDGE}, as well as the sensitivities of
LIGO, LISA and the Einstein Telescope (ET), a next-generation terrestrial laser interferometer~\cite{Badurina:2021rgt}. Also shown are the calculated GW signals
from the mergers of BH pairs with total masses 60, $10^4$ and $10^7$ solar masses at redshifts from 0.1 to 10.
As can be seen in the right panel of Fig.~\ref{fig:GWs}, 
the atom interferometers would have unique potential for measuring the mergers of IMBH pairs weighing beteen $10^2$ and
$10^5$ solar masses if they occur with redshifts $\lesssim 10$. The left panel shows that they could also observe the early infall stages of
BH mergers whose final stages could be detected by LIGO or ET, and their observations could be used to predict the
times and directions of these subsequent mergers, providing alerts that could trigger subsequent multi-messenger
observations. Conversely, inspiral measurements by LISA could be used to make predictions for subsequent AION or AEDGE
mergers.

The steep dashed lines in the left panel of Fig.~\ref{fig:GWs} and in the central part of the right panel
are estimates of the possible GGN
background, which is important at frequencies below ${\cal O}(1)$~Hz in terrestrial atom interferometers
and will require mitigation if their full GW potential is to be realised. There is no GGN background for
space-borne experiments, but there are backgrounds from unresolved galactic and extragalactic binaries
at frequencies $\lesssim {\cal O}(10^{-2})$~Hz that impact the prospective sensitivities of LISA and AEDGE.
The dashed lines at low frequencies $\lesssim {\cal O}(10^{-3})$~Hz in the left panel of Fig.~\ref{fig:GWs}
and large masses and redshifts in the right panel indicate the possible gain in sensitivity
for AEDGE that could be obtained by operating with atom clouds outside the spacecraft at distances
${\cal O}(10^{2})$~m.

\begin{figure}
\centering 
\includegraphics[width=6.25cm]{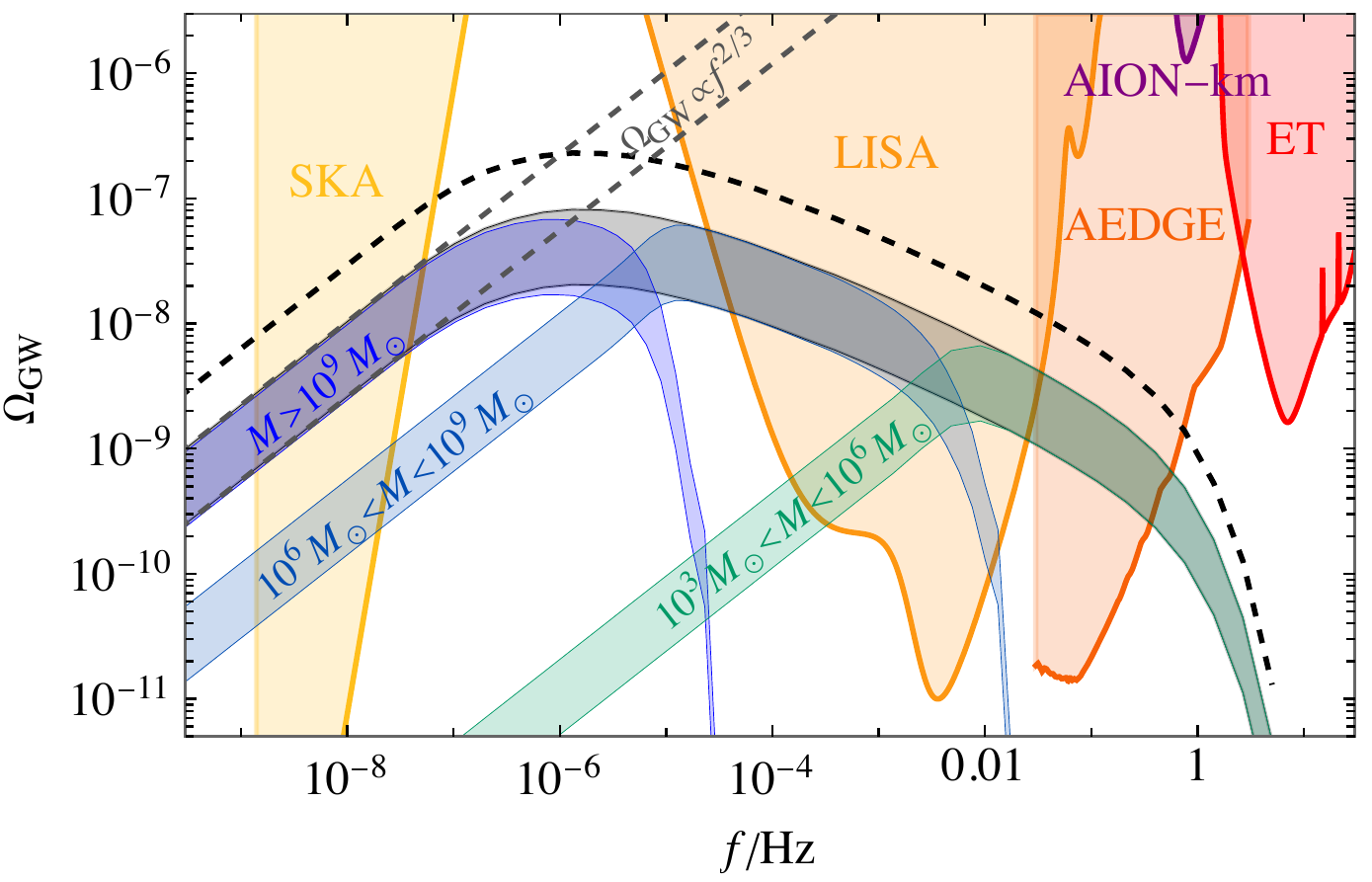}
\includegraphics[width=7.75cm]{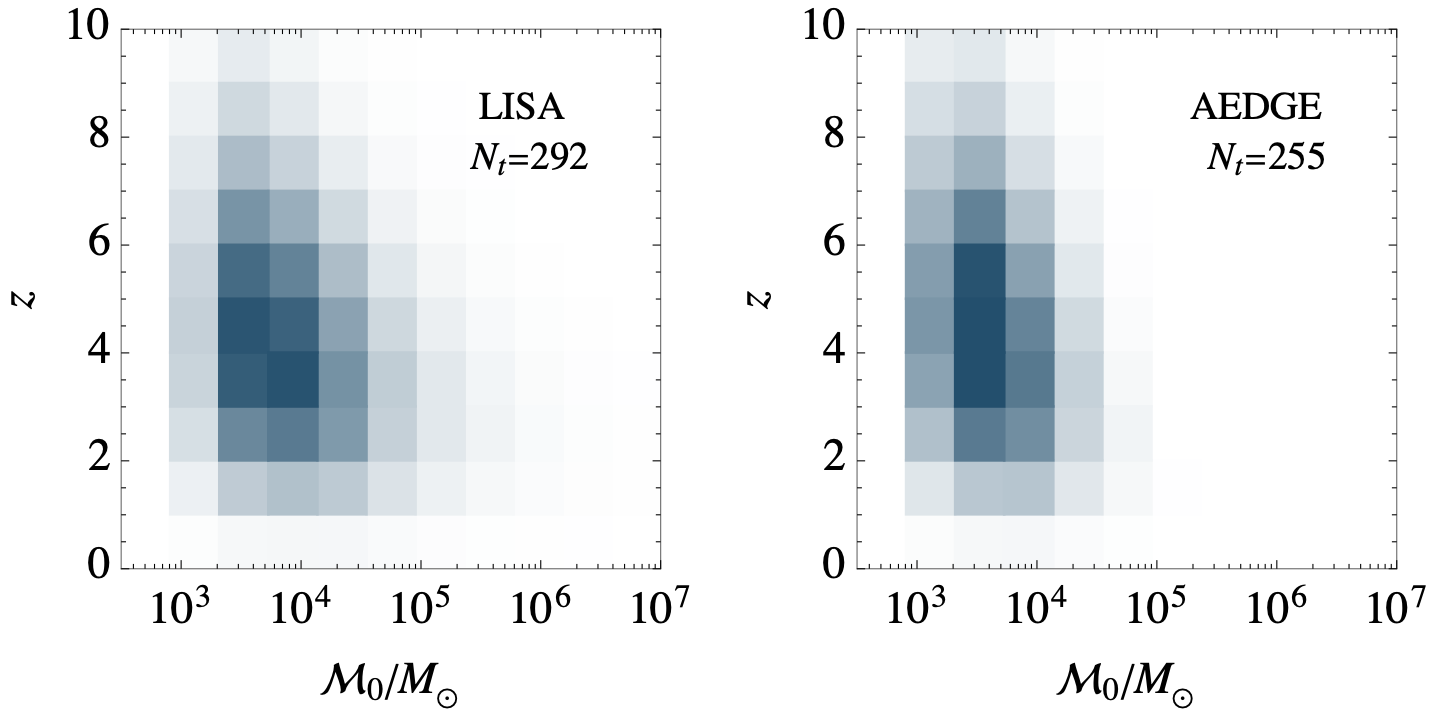}
\includegraphics[width=1cm]{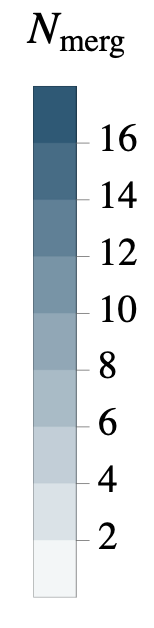}
\caption{\it Left panel: The mean GW energy density spectrum from massive BH mergers compared with the sensitivities of different experiments. The colored bands show the spectra from SMBHs and IMBHs in the indicated mass ranges, assuming a universal merger efficiency factor $p_{\rm BH} = 0.17^{+0.18}_{-0.08}$, whereas the black dashed curve shows the case where $p_{\rm BH} = 1$. The shaded regions show the prospective sensitivities of SKA~\cite{Janssen:2014dka}, LISA~\cite{Audley:2017drz}, AEDGE~\cite{AEDGE,Badurina:2021rgt}, AION-km~\cite{AION,Badurina:2021rgt} and ET~\cite{Sathyaprakash:2012jk}. Right panel: The expected numbers of detectable GW events generated during the last day before the merger in a year of observation by LISA and AEDGE as functions of the chirp mass $\mathcal{M}$ and redshift $z$, assuming a universal merger efficiency factor of $p_{\rm BH}=0.17$.
Figure adapted from~\cite{Ellis:2023owy}.}
\label{fig:EFHRUVV}
\end{figure}

It is all very well to be sensitive in principle to IMBH mergers, but how many such events could be expected? This question was addressed in a simple calculation using the extended Press-Schechter model to calculate the galactic halo mass function and merger rate, combined with a
simple relation between the halo masses and those of their central black holes and a universal merger efficiency factor $p_{\rm BH}$~\cite{Ellis:2023owy,Ellis:2023dgf}. The left panel of Fig.~\ref{fig:EFHRUVV} shows the mean GW energy density predicted in this model as a function of frequency, assuming a universal value of $p_{\rm BH} = 0.17^{+0.18}_{-0.08}$ suggested by International Pulsar Timing Array data~\cite{Ellis:2023owy} (recent NANOGrav data~\cite{NANOGrav:2023gor} suggest a larger value of $p_{\rm BH}$~\cite{Ellis:2023dgf}). Below a frequency $f \sim 10$~nHz most of the GWs are due to unresolved sources, but these become distinguishable at higher frequencies. The right panel of Fig.~\ref{fig:EFHRUVV} compares the expected numbers of detectable GW events generated during the last day before the merger in a year of observation by LISA and AEDGE as functions of the chirp mass $\mathcal{M}$ and redshift $z$, assuming a universal merger efficiency factor of $p_{\rm BH}=0.17$. AEDGE could see an interesting number of IMBH mergers, and many of these would have been detectable (and hence predictable) by LISA during the previous year, illustrating the synergy between these two detectors.

\begin{figure}
\centering 
\includegraphics[width=6cm]{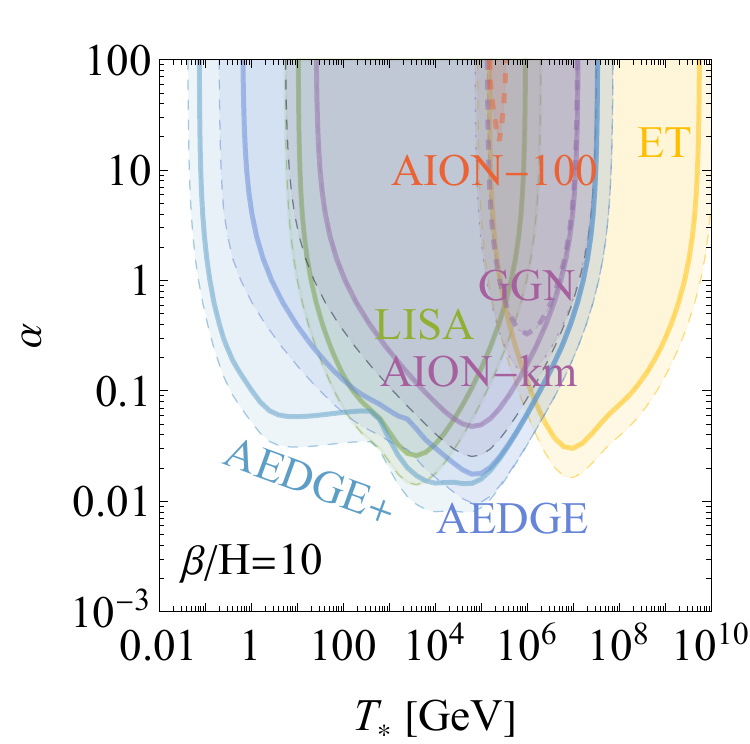}
\includegraphics[width=9cm]{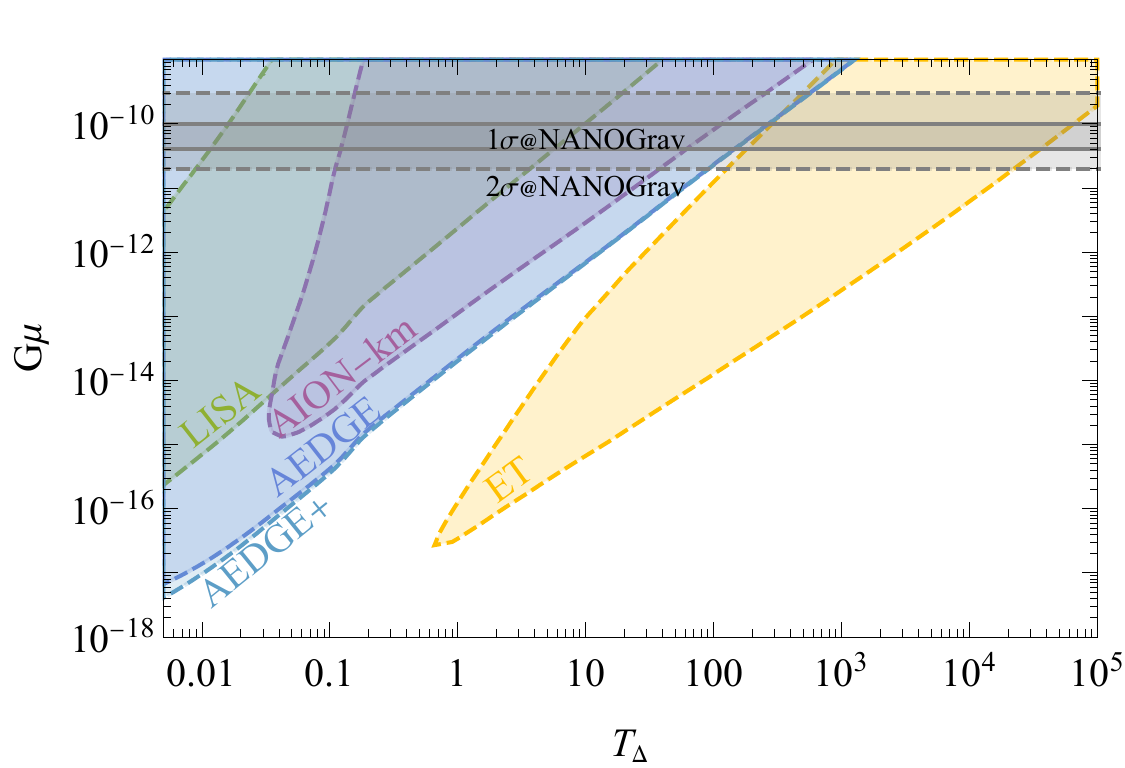}
\caption{\it Left panel: Sensitivities of AION-100 and -km, AEDGE and AEDGE+
and LISA in the $(T_*, \alpha)$ plane
to GWs from generic first-order transitions with the transition rate $\beta/H = 10$.
The thick solid lines are for $SNR=10$ and the dashed lines for $SNR=1$.
 Right panel: Sensitivities of different detectors
 to a 10\% modification of the string GW signal
 due to a change in the cosmological expansion rate at a temperature $T_\Delta$,
 as a function of the string tension $G\mu$. From~\cite{Badurina:2021rgt}.}
\label{fig:earlyuniverse}
\end{figure}

Atom interferometers are also sensitive to possible
GWs from fundamental physics sources in the early
Universe. The left panel of Fig.~\ref{fig:earlyuniverse} compares the sensitivities of different detectors to GWs
from generic first-order phase transitions with the
indicated transition temperature $T_*$, strength $\alpha$
and transition rate $\beta/H = 10$. Other possible
early Universe sources of GWs are networks of cosmic strings. The right panel of  Fig.~\ref{fig:earlyuniverse}
compares the sensitivities of different detectors to a 10\%
modification of the strength of the GW signal generated
by cosmic strings with tension $G \mu$ due to a change in the
cosmic expansion rate at a temperature $T_\Delta$. 
We see that AION and AEDGE complement well the sensitivities of LISA and ET~\cite{Badurina:2021rgt}: see also an analysis of cosmic strings~\cite{Ellis:2023tsl} in light of recent NANOGrav data~\cite{NANOGrav:2023gor}.

\subsection{Other Probes of Fundamental Physics}
\label{others}

In addition to the headline topics of the search for ULDM and measurements of mid-frequency
GWs, atom interferometers can probe fundamental physics in a number of other ways, of which we now
mention a few.

\begin{figure}
\centering
\includegraphics[width=0.7\textwidth]{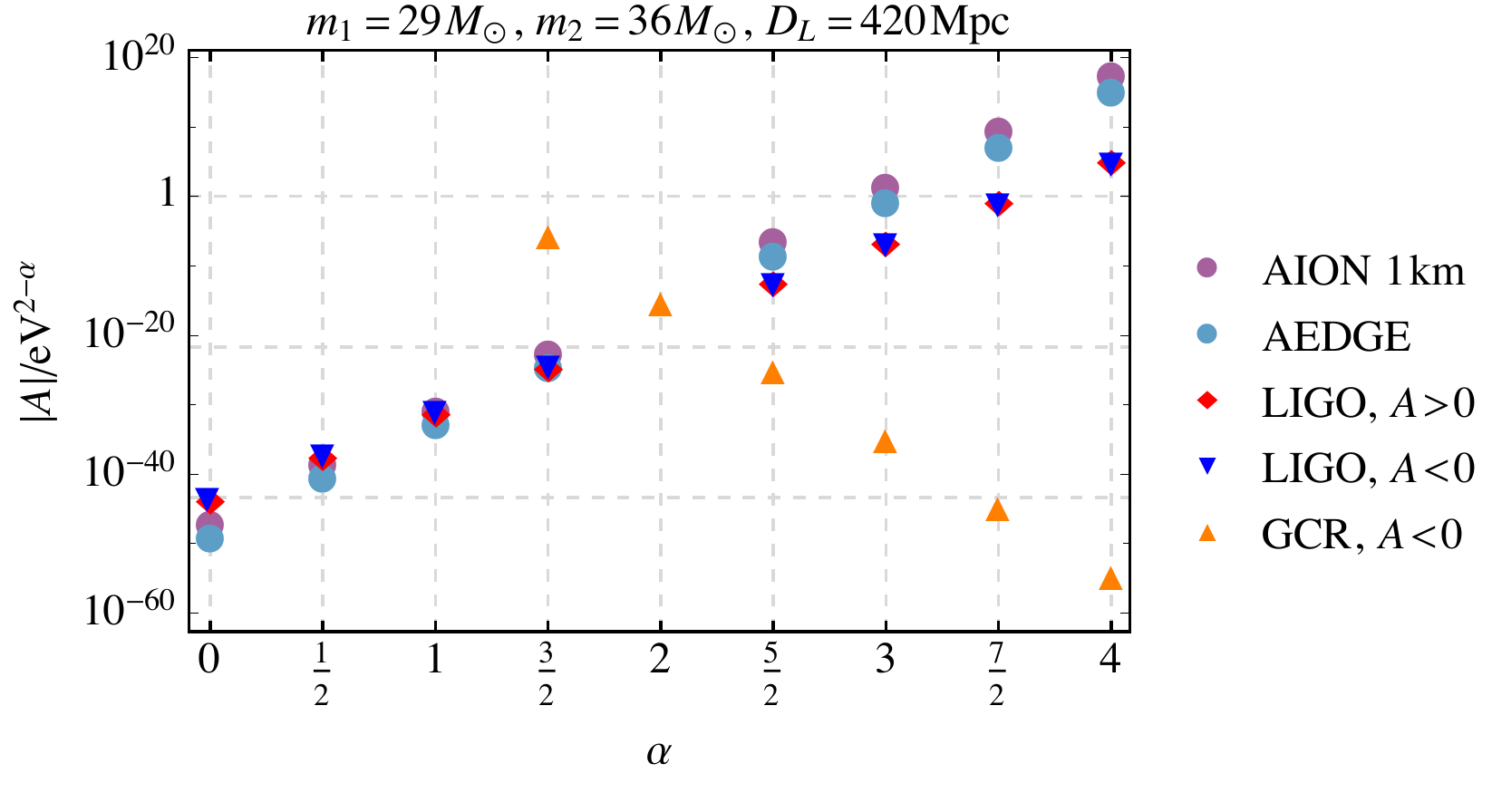}
\caption{\it 90\% CL upper bounds for the magnitude of the modification~\eqref{eq:dispersion} to the relativistic dispersion relation for GWs for different values of $\alpha$. The LIGO constraints are taken from Ref.~\cite{LIGOScientific:2019fpa}, the gravitational Cherenkov radiation (GCR) constraints are taken from Ref.~\cite{Kiyota:2015dla}, and the AION and AEDGE sensitivities are from~\cite{Ellis:2020lxl}.}
\label{fig:Aalpha}
\end{figure}

One may consider simple modifications of the dispersion relation for propagating modes of GWs
of the general form
\begin{equation} 
E^2 = p^2 + A p^\alpha \,,
\label{eq:dispersion}
\end{equation}
where $E$ is the energy and $p$ the momentum of the mode. 
This causes frequency-dependent phase shifts of the GWs from BH binary sources,
which are constrained by experimental data. Fig.~\ref{fig:Aalpha} compares constraints from LIGO data and
gravitational Cherenkov radiation with the potential sensitivities of measurements of a `LIGO-like'
BH merger at a distance $D_L = 420$~Mpc by AION~1km and AEDGE
to the possible value of the parameter $A$ in (\ref{eq:dispersion}) for different values of $\alpha$.
We see that AION~1km and AEDGE provide the strongest constraints for $\alpha \le 1$, and note that
the case $\alpha = 0$ corresponds to a graviton mass $m = \sqrt{A}$. Data from LIGO have established~\cite{LIGOScientific:2021sio} the
upper limit
\begin{equation}
m \; < 1.27 \times 10^{-23} \; {\rm eV} \, .
\label{LIGOlimit}
\end{equation}
The longer observing time with AION~1km of an event similar to the LIGO/Virgo discovery event would improve the sensitivity to $m < 1.1 \times 10^{-24}$~ eV at the 90\% CL, observations with AEDGE for 60 days before the binary merger would be sensitive to $m <1.3 \times 10^{-25}$~eV at the 90\% CL, and AEDGE observations of the infall stages of a BH binary with total mass $10^4$ solar masses at redshift $z = 1$ would be sensitive to
\begin{equation}
m \; < 8.1 \times 10^{-27} \; {\rm eV} 
\end{equation}
at the 90\% CL, an improvement on (\ref{LIGOlimit}) by over three orders of magnitude. As concerns Lorentz violation, AEDGE measurements could improve on the LIGO sensitivity by ${\cal O}(1000)$ for $\alpha = 1/2$ and by ${\cal O}(10)$ for $\alpha = 1$~\cite{Ellis:2020lxl}.

The Einstein Equivalence Principle (EEP) may be probed by measuring the relative accelerations of two different species in an atom interferometer.
A pioneering experiment of this type at Stanford used wave packets of $^{85}$Rb and $^{87}$Rb falling freely in the Earth’s gravitational field for 2 seconds and comparing their interference fringes. This experiment measured the
E{\" o}tv{\" o}s parameter $\eta = [1.6 \pm 1.8 ({\rm stat.}) \pm 3.4 ({\rm sys.})] \times 10^{-12}$, consistent with zero~\cite{Asenbaum:2020era}. This is one of the most sensitive
tests of the EEP in a terrestrial experiment, and future large atom interferometers may have interesting capabilities to test the EEP. For comparison, the space-borne
MICROSCOPE experiment comparing titanium and platinum alloys in free fall for five months used differential electrostatic accelerometers to measure $\eta ({\rm Ti, Pt}) = [-1.5 \pm 2.3 ({\rm stat.}) \pm 1.5 ({\rm syst.})] \times 10^{-15}$~\cite{MICROSCOPE:2022doy} and the STE-QUEST
proposal for a cold atom experiment in space aims to measure $\eta ({\rm Rb, K})$
with a precision $\sim 10^{-17}$~\cite{Ahlers:2022jdt}.

Another possibility for future large atom interferometers is to study the gravitational Aharonov-Bohm effect, following the pioneering Stanford experiment mentioned earlier that made a first measurement of this effect by splitting an atom cloud into two wave packets, one of
which passed close to a large source mass whereas the other was distant~\cite{Overstreet:2021hea}. In addition
to the classical effect of the gravitational field on the trajectory of the
first wave packet, the atom interferometer experiment was able to measure a 
quantum-mechanical phase shift due to the gravitational potential difference
between the two trajectories. Considering this
experiment in a quantum superposition of reference frames, 
it has been suggested~\cite{Overstreet:2022zgq}
that this result could be interpreted as providing evidence for the quantum superposition of different gravitational field configurations.

These few examples illustrate the new possibilities for probing
fundamental physics that are opened up by atom interferometers.

\section{Summary and Conclusion}

Large-scale atom interferometry is an emerging field that has the potential to contribute greatly to our understanding of the universe. In this article we have reviewed the basic principles of such devices, which exploit the wave-like nature of atoms by splitting a beam of atoms into clouds following two distinct trajectories before then recombining them to create an interference pattern that can be used to extract even tiny phase shifts.
Laboratory atom interferometry experiments have already provided one of the most sensitive terrestrial tests of Einstein's Equivalence Principle and demonstrated a gravitational analogue of the Aharanov-Bohm effect.

This review has highlighted how current atom interferometer projects can be scaled up from the baselines of ${\cal O}(10)$~m used in current experiments to baselines that are ${\cal O}(1)$~km on Earth, and potentially thousands of km in space. These large-scale atom interferometers have the potential to improve significantly our ability to detect and study ultralight dark matter and gravitational waves, which are some of the most elusive phenomena in modern physics.

For example, they can be used to search for ultralight bosonic dark matter that interacts with Standard Model particles by measuring its effects on the energy levels of the atoms. Also, they are sensitive to the passage of gravitational waves through the phase shifts induced by the distortions of space-time that they generate.
Besides these headline fundamental physics objectives, large-scale atom interferometers may also provide additional tests of the Equivalence Principle and more sensitive measurements of gravitational phase shifts. They could also be used to look for new fundamental interactions and probe the validity of quantum mechanics.

A recent workshop hosted by CERN brought together members of the cold atom, particle physics, astrophysics and cosmology communities to discuss the fundamental scientific objectives of terrestrial long-baseline atom interferometers, review ongoing projects, and consider the possibilities for future projects at the km scale~\cite{TVLBAI}. Based on these discussions, work is underway to develop a roadmap for international efforts to construct one or more km-scale atom interferometers in the 2030s.

In parallel to these exciting prospects in fundamental physics research, atom interferometry has many potential practical applications. For example, it can be used in highly sensitive inertial sensors suitable for precision measurements of use in navigation and geophysical exploration. It also enables high-precision measurements of the local gravitational acceleration that are useful for detecting underground structures and potentially Earth Observation from space that can contribute to understanding the impact of climate change. 
Prospects for such measurements and their synergies with atom interferometer experiments in fundamental physics have been reviewed in a previous workshop that proposed a roadmap for cold atoms in space, with milestones paving the way for a possible space-borne atom interferometer experiment to undertake searches for ultralight dark matter and intermediate-frequency gravitational waves in the 2040s~\cite{Alonso:2022oot}.

In conclusion, large-scale atom interferometry is an exciting field with many potential applications in fundamental physics, with the potential for important discoveries. As cold-atom technology continues to advance and new projects are developed, we can expect it to enable breakthroughs in our understanding of fundamental physics and the Universe as a whole. Moreover, atom interferometry has the potential to revolutionise many scientific fields beyond fundamental physics research.

\section*{Acknowledgements}

This work was supported by UKRI through its Quantum Technology for
Fundamental Physics programme, via the following grants from EPSRC and STFC
in the framework of the AION Consortium:
STFC Grants ST/T006579/1 and ST/W006200/1 to the University of Cambridge,
STFC Grant ST/T006994/1 to Imperial College London, and
STFC Grant ST/T00679X/1 to King's College London.
For the purpose of open access, the authors have applied a Creative Commons Attribution (CC-BY) licence to any Author Accepted Manuscript version arising from this submission.

\bibliographystyle{JHEP}
\bibliography{main}

\providecommand{\href}[2]{#2}\begingroup\raggedright\begin{thebibliography}{10}

\bibitem{Zwicky:1933gu}
F.~Zwicky, \emph{{Die Rotverschiebung von extragalaktischen Nebeln}},
  \href{https://doi.org/10.1007/s10714-008-0707-4}{\emph{Helv. Phys. Acta}
  {\bfseries 6} (1933) 110}.

\bibitem{Zwicky:1937zza}
F.~Zwicky, \emph{{On the Masses of Nebulae and of Clusters of Nebulae}},
  \href{https://doi.org/10.1086/143864}{\emph{Astrophys. J.} {\bfseries 86}
  (1937) 217}.

\bibitem{Rubin:1970zza}
V.C.~Rubin and W.K.~Ford, Jr., \emph{{Rotation of the Andromeda Nebula from a
  Spectroscopic Survey of Emission Regions}},
  \href{https://doi.org/10.1086/150317}{\emph{Astrophys. J.} {\bfseries 159}
  (1970) 379}.

\bibitem{Bertone:2016nfn}
G.~Bertone and D.~Hooper, \emph{{History of dark matter}},
  \href{https://doi.org/10.1103/RevModPhys.90.045002}{\emph{Rev. Mod. Phys.}
  {\bfseries 90} (2018) 045002}
  [\href{https://arxiv.org/abs/1605.04909}{{\ttfamily 1605.04909}}].

\bibitem{Carr:2021bzv}
B.~Carr and F.~Kuhnel, \emph{{Primordial black holes as dark matter
  candidates}},
  \href{https://doi.org/10.21468/SciPostPhysLectNotes.48}{\emph{SciPost Phys.
  Lect. Notes} {\bfseries 48} (2022) 1}
  [\href{https://arxiv.org/abs/2110.02821}{{\ttfamily 2110.02821}}].

\bibitem{ATLAS:2022ihe}
{\scshape ATLAS} collaboration, \emph{{Search for supersymmetry in final states
  with missing transverse momentum and three or more $b$-jets in 139 fb$^{-1}$
  of proton$-$proton collisions at $\sqrt{s} = 13$ TeV with the ATLAS
  detector}},  \href{https://arxiv.org/abs/2211.08028}{{\ttfamily 2211.08028}}.

\bibitem{CMS:2023ktc}
{\scshape CMS} collaboration, \emph{{Search for top squarks in the four-body
  decay mode with single lepton final states in proton-proton collisions at
  $\sqrt{s}$ = 13 TeV}},  \href{https://arxiv.org/abs/2301.08096}{{\ttfamily
  2301.08096}}.

\bibitem{LZ:2022ufs}
{\scshape LZ} collaboration, \emph{{First Dark Matter Search Results from the
  LUX-ZEPLIN (LZ) Experiment}},
  \href{https://arxiv.org/abs/2207.03764}{{\ttfamily 2207.03764}}.

\bibitem{Graham:2015ifn}
P.W.~Graham, D.E.~Kaplan, J.~Mardon, S.~Rajendran and W.A.~Terrano, \emph{{Dark
  Matter Direct Detection with Accelerometers}},
  \href{https://doi.org/10.1103/PhysRevD.93.075029}{\emph{Phys. Rev. D}
  {\bfseries 93} (2016) 075029}
  [\href{https://arxiv.org/abs/1512.06165}{{\ttfamily 1512.06165}}].

\bibitem{Graham:2012sy}
P.W.~Graham, J.M.~Hogan, M.A.~Kasevich and S.~Rajendran, \emph{{A New Method
  for Gravitational Wave Detection with Atomic Sensors}},
  \href{https://doi.org/10.1103/PhysRevLett.110.171102}{\emph{Phys. Rev. Lett.}
  {\bfseries 110} (2013) 171102}
  [\href{https://arxiv.org/abs/1206.0818}{{\ttfamily 1206.0818}}].

\bibitem{LIGOScientific:2016aoc}
{\scshape LIGO Scientific, Virgo} collaboration, \emph{{Observation of
  Gravitational Waves from a Binary Black Hole Merger}},
  \href{https://doi.org/10.1103/PhysRevLett.116.061102}{\emph{Phys. Rev. Lett.}
  {\bfseries 116} (2016) 061102}
  [\href{https://arxiv.org/abs/1602.03837}{{\ttfamily 1602.03837}}].

\bibitem{LIGOScientific:2017vwq}
{\scshape LIGO Scientific, Virgo} collaboration, \emph{{GW170817: Observation
  of Gravitational Waves from a Binary Neutron Star Inspiral}},
  \href{https://doi.org/10.1103/PhysRevLett.119.161101}{\emph{Phys. Rev. Lett.}
  {\bfseries 119} (2017) 161101}
  [\href{https://arxiv.org/abs/1710.05832}{{\ttfamily 1710.05832}}].

\bibitem{EventHorizonTelescope:2019dse}
{\scshape Event Horizon Telescope} collaboration, \emph{{First M87 Event
  Horizon Telescope Results. I. The Shadow of the Supermassive Black Hole}},
  \href{https://doi.org/10.3847/2041-8213/ab0ec7}{\emph{Astrophys. J. Lett.}
  {\bfseries 875} (2019) L1}
  [\href{https://arxiv.org/abs/1906.11238}{{\ttfamily 1906.11238}}].

\bibitem{EventHorizonTelescope:2022wkp}
{\scshape Event Horizon Telescope} collaboration, \emph{{First Sagittarius A*
  Event Horizon Telescope Results. I. The Shadow of the Supermassive Black Hole
  in the Center of the Milky Way}},
  \href{https://doi.org/10.3847/2041-8213/ac6674}{\emph{Astrophys. J. Lett.}
  {\bfseries 930} (2022) L12}.

\bibitem{Audley:2017drz}
{\scshape LISA} collaboration, \emph{{Laser Interferometer Space Antenna}},
  \href{https://arxiv.org/abs/1702.00786}{{\ttfamily 1702.00786}}.

\bibitem{Greene:2019vlv}
J.E.~Greene, J.~Strader and L.C.~Ho, \emph{{Intermediate-Mass Black Holes}},
  \href{https://doi.org/10.1146/annurev-astro-032620-021835}{\emph{Ann. Rev.
  Astron. Astrophys.} {\bfseries 58} (2020) 257}
  [\href{https://arxiv.org/abs/1911.09678}{{\ttfamily 1911.09678}}].

\bibitem{Badurina:2021rgt}
L.~Badurina, O.~Buchmueller, J.~Ellis, M.~Lewicki, C.~McCabe and V.~Vaskonen,
  \emph{{Prospective sensitivities of atom interferometers to gravitational
  waves and ultralight dark matter}},
  \href{https://doi.org/10.1098/rsta.2021.0060}{\emph{Phil. Trans. A. Math.
  Phys. Eng. Sci.} {\bfseries 380} (2021) 20210060}
  [\href{https://arxiv.org/abs/2108.02468}{{\ttfamily 2108.02468}}].

\bibitem{Asenbaum:2020era}
P.~Asenbaum, C.~Overstreet, M.~Kim, J.~Curti and M.A.~Kasevich,
  \emph{{Atom-Interferometric Test of the Equivalence Principle at the
  $10^{-12}$ Level}},
  \href{https://doi.org/10.1103/PhysRevLett.125.191101}{\emph{Phys. Rev. Lett.}
  {\bfseries 125} (2020) 191101}
  [\href{https://arxiv.org/abs/2005.11624}{{\ttfamily 2005.11624}}].

\bibitem{BILARDELLO2016764}
M.~Bilardello, S.~Donadi, A.~Vinante and A.~Bassi, \emph{Bounds on collapse
  models from cold-atom experiments},
  \href{https://doi.org/https://doi.org/10.1016/j.physa.2016.06.134}{\emph{Physica
  A: Statistical Mechanics and its Applications} {\bfseries 462} (2016) 764}.

\bibitem{Overstreet:2021hea}
C.~Overstreet, P.~Asenbaum, J.~Curti, M.~Kim and M.A.~Kasevich,
  \emph{{Observation of a gravitational Aharonov-Bohm effect}},
  \href{https://doi.org/10.1126/science.abl7152}{\emph{Science} {\bfseries 375}
  (2021) abl7152}.

\bibitem{Stray2022}
B.~Stray, A.~Lamb, A.~Kaushik, J.~Vovrosh, A.~Rodgers, J.~Winch et~al.,
  \emph{{Quantum sensing for gravity cartography}},
  \href{https://doi.org/10.1038/s41586-021-04315-3}{\emph{Nature 2022 602:7898}
  {\bfseries 602} (2022) 590}.

\bibitem{deBroglie1924}
L.~de~Broglie, \emph{Recherches sur la théorie des quanta}, {\emph{Annales de
  Physique} {\bfseries 10} (1924) 22}.

\bibitem{young1804experiments}
T.~Young, \emph{Experiments and calculations relative to physical optics},
  {\emph{Philosophical Transactions of the Royal Society of London} {\bfseries
  94} (1804) 1}.

\bibitem{Davisson1927}
C.~Davisson and L.H.~Germer, \emph{Diffraction of electrons by a crystal of
  nickel}, \href{https://doi.org/10.1103/PhysRev.30.705}{\emph{Phys. Rev.}
  {\bfseries 30} (1927) 705}.

\bibitem{Thomson1927}
G.P.~Thomson and A.~Reid, \emph{{Diffraction of Cathode Rays by a Thin Film}},
  \href{https://doi.org/10.1038/119890a0}{\emph{Nature 1927 119:3007}
  {\bfseries 119} (1927) 890}.

\bibitem{Estermann1930}
I.~Estermann and O.~Stern, \emph{{Beugung von Molekularstrahlen}},
  \href{https://doi.org/10.1007/BF01340293/METRICS}{\emph{Zeitschrift f{\"{u}}r
  Physik} {\bfseries 61} (1930) 95}.

\bibitem{Anderson:1995gf}
M.H.~Anderson, J.R.~Ensher, M.R.~Matthews, C.E.~Wieman and E.A.~Cornell,
  \emph{{Observation of Bose-Einstein condensation in a dilute atomic vapor}},
  \href{https://doi.org/10.1126/science.269.5221.198}{\emph{Science} {\bfseries
  269} (1995) 198}.

\bibitem{Davis95}
K.B.~Davis, M.O.~Mewes, M.R.~Andrews, N.J.~van Druten, D.S.~Durfee, D.M.~Kurn
  et~al., \emph{Bose-{E}instein condensation in a gas of sodium atoms},
  \href{https://doi.org/10.1103/PhysRevLett.75.3969}{\emph{Phys. Rev. Lett.}
  {\bfseries 75} (1995) 3969}.

\bibitem{Carnal1991}
O.~Carnal and J.~Mlynek, \emph{Young's double-slit experiment with atoms: A
  simple atom interferometer},
  \href{https://doi.org/10.1103/PhysRevLett.66.2689}{\emph{Phys. Rev. Lett.}
  {\bfseries 66} (1991) 2689}.

\bibitem{Keith1991}
D.W.~Keith, C.R.~Ekstrom, Q.A.~Turchette and D.E.~Pritchard, \emph{An
  interferometer for atoms},
  \href{https://doi.org/10.1103/PhysRevLett.66.2693}{\emph{Phys. Rev. Lett.}
  {\bfseries 66} (1991) 2693}.

\bibitem{Riehle1991}
F.~Riehle, T.~Kisters, A.~Witte, J.~Helmcke and C.J.~Bord\'e, \emph{Optical
  {R}amsey spectroscopy in a rotating frame: Sagnac effect in a matter-wave
  interferometer},
  \href{https://doi.org/10.1103/PhysRevLett.67.177}{\emph{Phys. Rev. Lett.}
  {\bfseries 67} (1991) 177}.

\bibitem{kasevich91corr}
M.~Kasevich and S.~Chu, \emph{Atomic interferometry using stimulated raman
  transitions}, \href{https://doi.org/10.1103/PhysRevLett.67.181}{\emph{Phys.
  Rev. Lett.} {\bfseries 67} (1991) 181}.

\bibitem{Cronin2009}
A.D.~Cronin, J.~Schmiedmayer and D.E.~Pritchard, \emph{Optics and
  interferometry with atoms and molecules},
  \href{https://doi.org/10.1103/RevModPhys.81.1051}{\emph{Rev. Mod. Phys.}
  {\bfseries 81} (2009) 1051}.

\bibitem{zehnder1891neuer}
L.~Zehnder, \emph{Ein neuer {I}nterferenzrefraktor}, {\emph{Zeitschrift f{\"u}r
  Instrumentenkunde} {\bfseries 11} (1891) 275}.

\bibitem{mach1892ueber}
L.~Mach, \emph{Ueber einen {I}nterferenzrefraktor}, {\emph{Zeitschrift f{\"u}r
  Instrumentenkunde} {\bfseries 12} (1892) 89}.

\bibitem{Ludlow2015}
A.D.~Ludlow, M.M.~Boyd, J.~Ye, E.~Peik and P.O.~Schmidt, \emph{Optical atomic
  clocks}, \href{https://doi.org/10.1103/RevModPhys.87.637}{\emph{Rev. Mod.
  Phys.} {\bfseries 87} (2015) 637}.

\bibitem{rasel1995atom}
E.M.~Rasel, M.K.~Oberthaler, H.~Batelaan, J.~Schmiedmayer and A.~Zeilinger,
  \emph{Atom wave interferometry with diffraction gratings of light},
  {\emph{Physical Review Letters} {\bfseries 75} (1995) 2633}.

\bibitem{Kialka2022}
F.~Kiałka, Y.Y.~Fein, S.~Pedalino, S.~Gerlich and M.~Arndt, \emph{A roadmap
  for universal high-mass matter-wave interferometry},
  \href{https://doi.org/10.1116/5.0080940}{\emph{AVS Quantum Science}
  {\bfseries 4} (2022) 020502}
  [\href{https://arxiv.org/abs/https://doi.org/10.1116/5.0080940}{{\ttfamily
  https://doi.org/10.1116/5.0080940}}].

\bibitem{Rudolph:2019vcv}
J.~Rudolph, T.~Wilkason, M.~Nantel, H.~Swan, C.M.~Holland, Y.~Jiang et~al.,
  \emph{{Large Momentum Transfer Clock Atom Interferometry on the 689 nm
  Intercombination Line of Strontium}},
  \href{https://doi.org/10.1103/PhysRevLett.124.083604}{\emph{Phys. Rev. Lett.}
  {\bfseries 124} (2020) 083604}
  [\href{https://arxiv.org/abs/1910.05459}{{\ttfamily 1910.05459}}].

\bibitem{Yu2011}
N.~Yu and M.~Tinto, \emph{Gravitational wave detection with single-laser atom
  interferometers},
  \href{https://doi.org/10.1007/s10714-010-1055-8}{\emph{General Relativity and
  Gravitation} {\bfseries 43} (2011) 1943}.

\bibitem{Badurina:2022ngn}
L.~Badurina, V.~Gibson, C.~McCabe and J.~Mitchell, \emph{{Ultralight dark
  matter searches at the sub-Hz frontier with atom multigradiometry}},
  \href{https://doi.org/10.1103/PhysRevD.107.055002}{\emph{Phys. Rev. D}
  {\bfseries 107} (2023) 055002}
  [\href{https://arxiv.org/abs/2211.01854}{{\ttfamily 2211.01854}}].

\bibitem{wilkason2022atom}
T.~Wilkason, M.~Nantel, J.~Rudolph, Y.~Jiang, B.E.~Garber, H.~Swan et~al.,
  \emph{Atom interferometry with floquet atom optics}, {\emph{Physical Review
  Letters} {\bfseries 129} (2022) 183202}.

\bibitem{Muniz2021}
J.A.~Muniz, D.J.~Young, J.R.K.~Cline and J.K.~Thompson, \emph{Cavity-qed
  measurements of the $^{87}\mathrm{Sr}$ millihertz optical clock transition
  and determination of its natural linewidth},
  \href{https://doi.org/10.1103/PhysRevResearch.3.023152}{\emph{Phys. Rev.
  Res.} {\bfseries 3} (2021) 023152}.

\bibitem{hosten2016measurement}
O.~Hosten, N.J.~Engelsen, R.~Krishnakumar and M.A.~Kasevich, \emph{Measurement
  noise 100 times lower than the quantum-projection limit using entangled
  atoms}, {\emph{Nature} {\bfseries 529} (2016) 505}.

\bibitem{Damour:2010rm}
T.~Damour and J.F.~Donoghue, \emph{{Phenomenology of the Equivalence Principle
  with Light Scalars}},
  \href{https://doi.org/10.1088/0264-9381/27/20/202001}{\emph{Class. Quant.
  Grav.} {\bfseries 27} (2010) 202001}
  [\href{https://arxiv.org/abs/1007.2790}{{\ttfamily 1007.2790}}].

\bibitem{Damour:2010rp}
T.~Damour and J.F.~Donoghue, \emph{{Equivalence Principle Violations and
  Couplings of a Light Dilaton}},
  \href{https://doi.org/10.1103/PhysRevD.82.084033}{\emph{Phys. Rev.}
  {\bfseries D82} (2010) 084033}
  [\href{https://arxiv.org/abs/1007.2792}{{\ttfamily 1007.2792}}].

\bibitem{MAGIS-100:2021etm}
{\scshape MAGIS-100} collaboration, \emph{{Matter-wave Atomic Gradiometer
  Interferometric Sensor (MAGIS-100)}},
  \href{https://doi.org/10.1088/2058-9565/abf719}{\emph{Quantum Sci. Technol.}
  {\bfseries 6} (2021) 044003}
  [\href{https://arxiv.org/abs/2104.02835}{{\ttfamily 2104.02835}}].

\bibitem{AION}
L.~Badurina et~al., \emph{{AION: An Atom Interferometer Observatory and
  Network}}, \href{https://doi.org/10.1088/1475-7516/2020/05/011}{\emph{JCAP}
  {\bfseries 05} (2020) 011}
  [\href{https://arxiv.org/abs/1911.11755}{{\ttfamily 1911.11755}}].

\bibitem{Canuel:2017rrp}
B.~Canuel et~al., \emph{{Exploring gravity with the MIGA large scale atom
  interferometer}},
  \href{https://doi.org/10.1038/s41598-018-32165-z}{\emph{Sci. Rep.} {\bfseries
  8} (2018) 14064} [\href{https://arxiv.org/abs/1703.02490}{{\ttfamily
  1703.02490}}].

\bibitem{schlippert2020matter}
D.~Schlippert, C.~Meiners, R.~Rengelink, C.~Schubert, D.~Tell, {\'E}.~Wodey
  et~al., \emph{Matter-wave interferometry for inertial sensing and tests of
  fundamental physics},  in \emph{CPT AND LORENTZ SYMMETRY: Proceedings of the
  Eighth Meeting on CPT and Lorentz Symmetry}, pp.~37--40, World Scientific,
  2020.

\bibitem{Zhan:2019quq}
M.-S.~Zhan et~al., \emph{{ZAIGA: Zhaoshan Long-baseline Atom Interferometer
  Gravitation Antenna}},
  \href{https://doi.org/10.1142/S0218271819400054}{\emph{Int. J. Mod. Phys.}
  {\bfseries D28} (2019) 1940005}
  [\href{https://arxiv.org/abs/1903.09288}{{\ttfamily 1903.09288}}].

\bibitem{AEDGE}
{\scshape AEDGE} collaboration, \emph{{AEDGE: Atomic Experiment for Dark Matter
  and Gravity Exploration in Space}},
  \href{https://doi.org/10.1140/epjqt/s40507-020-0080-0}{\emph{EPJ Quant.
  Technol.} {\bfseries 7} (2020) 6}
  [\href{https://arxiv.org/abs/1908.00802}{{\ttfamily 1908.00802}}].

\bibitem{Asenbaum:2016djh}
P.~Asenbaum, C.~Overstreet, T.~Kovachy, D.D.~Brown, J.M.~Hogan and
  M.A.~Kasevich, \emph{{Phase Shift in an Atom Interferometer due to Spacetime
  Curvature across its Wave Function}},
  \href{https://doi.org/10.1103/PhysRevLett.118.183602}{\emph{Phys. Rev. Lett.}
  {\bfseries 118} (2017) 183602}
  [\href{https://arxiv.org/abs/1610.03832}{{\ttfamily 1610.03832}}].

\bibitem{heise2022sanford}
J.~Heise, \emph{{The {S}anford {U}nderground {R}esearch {F}acility}},
  \href{https://arxiv.org/abs/arXiv:2203.08293}{{\ttfamily arXiv:2203.08293}}.

\bibitem{Arduini:2023wce}
G.~Arduini et~al., \emph{{A Long-Baseline Atom Interferometer at CERN:
  Conceptual Feasibility Study}},
  \href{https://arxiv.org/abs/arXiv:2304.00614}{{\ttfamily arXiv:2304.00614}}.

\bibitem{AION:2023fpx}
{\scshape AION} collaboration, \emph{{Centralised Design and Production of the
  Ultra-High Vacuum and Laser-Stabilisation Systems for the AION Ultra-Cold
  Strontium Laboratories}},
  \href{https://arxiv.org/abs/arXiv:2305.20060}{{\ttfamily arXiv:2305.20060}}.

\bibitem{ELGAR}
B.~Canuel et~al., \emph{{ELGAR\textemdash{}a European Laboratory for
  Gravitation and Atom-interferometric Research}},
  \href{https://doi.org/10.1088/1361-6382/aba80e}{\emph{Class. Quant. Grav.}
  {\bfseries 37} (2020) 225017}
  [\href{https://arxiv.org/abs/1911.03701}{{\ttfamily 1911.03701}}].

\bibitem{TVLBAI}
``{Terrestrial Very-Long-Baseline Atom Interferometry Workshop}.''
  \url{https://indico.cern.ch/event/1208783/}.

\bibitem{Alonso:2022oot}
I.~Alonso et~al., \emph{{Cold atoms in space: community workshop summary and
  proposed road-map}},
  \href{https://doi.org/10.1140/epjqt/s40507-022-00147-w}{\emph{EPJ Quant.
  Technol.} {\bfseries 9} (2022) 30}
  [\href{https://arxiv.org/abs/2201.07789}{{\ttfamily 2201.07789}}].

\bibitem{Liu:2018kn}
L.~Liu, D.-S.~L{\"u}, W.-B.~Chen, T.~Li, Q.-Z.~Qu, B.~Wang et~al.,
  \emph{In-orbit operation of an atomic clock based on laser-cooled 87rb
  atoms}, \href{https://doi.org/10.1038/s41467-018-05219-z}{\emph{Nature
  Communications} {\bfseries 9} (2018) 2760}.

\bibitem{MAIUS}
D.~Becker, M.D.~Lachmann, S.T.~Seidel, H.~Ahlers, A.N.~Dinkelaker, J.~Grosse
  et~al., \emph{Space-borne {B}ose--{E}instein condensation for precision
  interferometry},
  \href{https://doi.org/10.1038/s41586-018-0605-1}{\emph{Nature} {\bfseries
  562} (2018) 391}.

\bibitem{CAL}
E.R.~Elliott, M.C.~Krutzik, J.R.~Williams, R.J.~Thompson and D.C.~Aveline,
  \emph{{NASA}'s {C}old {A}tom {L}ab ({CAL}): system development and ground
  test status}, \href{https://doi.org/10.1038/s41526-018-0049-9}{\emph{npj
  Microgravity} {\bfseries 4} (2018) 16}.

\bibitem{Lezius}
M.~Lezius et~al., \emph{Space-borne frequency comb metrology},
  \href{https://doi.org/10.1038/s41526-018-0049-9}{\emph{Optica} {\bfseries 3}
  (2016) 1381}.

\bibitem{Dinkelaker}
A.~Dinkelaker et~al., \emph{Space-borne frequency comb metrology},
  \href{https://doi.org/10.1364/AO.56.001388}{\emph{Appl. Opt.} {\bfseries 56}
  (2017) 1388}.

\bibitem{PhysRevApplied.11.054068}
K.~D\"oringshoff et~al., \emph{Iodine frequency reference on a sounding
  rocket}, \href{https://doi.org/10.1103/PhysRevApplied.11.054068}{\emph{Phys.
  Rev. Applied} {\bfseries 11} (2019) 054068}.

\bibitem{Tino2019}
G.M.~Tino et~al., \emph{{SAGE: A Proposal for a Space Atomic Gravity
  Explorer}}, {\emph{Eur. Phys. J. D} {\bfseries Topical Issue on Quantum
  Technologies for Gravitational Physics} (2019) In press}
  [\href{https://arxiv.org/abs/1907.03867}{{\ttfamily 1907.03867}}].

\bibitem{Ahlers:2022jdt}
H.~Ahlers et~al., \emph{{STE-QUEST: Space Time Explorer and QUantum Equivalence
  principle Space Test}},
  \href{https://arxiv.org/abs/arXiv:2211.15412}{{\ttfamily arXiv:2211.15412}}.

\bibitem{Arvanitaki:2016fyj}
A.~Arvanitaki, P.W.~Graham, J.M.~Hogan, S.~Rajendran and K.~Van~Tilburg,
  \emph{{Search for light scalar dark matter with atomic gravitational wave
  detectors}}, \href{https://doi.org/10.1103/PhysRevD.97.075020}{\emph{Phys.
  Rev.} {\bfseries D97} (2018) 075020}
  [\href{https://arxiv.org/abs/1606.04541}{{\ttfamily 1606.04541}}].

\bibitem{Stadnik:2014tta}
Y.V.~Stadnik and V.V.~Flambaum, \emph{{Searching for dark matter and variation
  of fundamental constants with laser and maser interferometry}},
  \href{https://doi.org/10.1103/PhysRevLett.114.161301}{\emph{Phys. Rev. Lett.}
  {\bfseries 114} (2015) 161301}
  [\href{https://arxiv.org/abs/1412.7801}{{\ttfamily 1412.7801}}].

\bibitem{Derevianko:2016vpm}
A.~Derevianko, \emph{{Detecting dark-matter waves with a network of
  precision-measurement tools}},
  \href{https://doi.org/10.1103/PhysRevA.97.042506}{\emph{Phys. Rev. A}
  {\bfseries 97} (2018) 042506}
  [\href{https://arxiv.org/abs/1605.09717}{{\ttfamily 1605.09717}}].

\bibitem{MICROSCOPE:2022doy}
{\scshape MICROSCOPE} collaboration, \emph{{MICROSCOPE Mission: Final Results
  of the Test of the Equivalence Principle}},
  \href{https://doi.org/10.1103/PhysRevLett.129.121102}{\emph{Phys. Rev. Lett.}
  {\bfseries 129} (2022) 121102}
  [\href{https://arxiv.org/abs/2209.15487}{{\ttfamily 2209.15487}}].

\bibitem{Wagner:2012ui}
T.A.~Wagner, S.~Schlamminger, J.H.~Gundlach and E.G.~Adelberger,
  \emph{{Torsion-balance tests of the weak equivalence principle}},
  \href{https://doi.org/10.1088/0264-9381/29/18/184002}{\emph{Class. Quant.
  Grav.} {\bfseries 29} (2012) 184002}
  [\href{https://arxiv.org/abs/1207.2442}{{\ttfamily 1207.2442}}].

\bibitem{Hees:2016gop}
A.~Hees, J.~Guena, M.~Abgrall, S.~Bize and P.~Wolf, \emph{{Searching for an
  oscillating massive scalar field as a dark matter candidate using atomic
  hyperfine frequency comparisons}},
  \href{https://doi.org/10.1103/PhysRevLett.117.061301}{\emph{Phys. Rev. Lett.}
  {\bfseries 117} (2016) 061301}
  [\href{https://arxiv.org/abs/1604.08514}{{\ttfamily 1604.08514}}].

\bibitem{Branca:2016rez}
A.~Branca et~al., \emph{{Search for an Ultralight Scalar Dark Matter Candidate
  with the AURIGA Detector}},
  \href{https://doi.org/10.1103/PhysRevLett.118.021302}{\emph{Phys. Rev. Lett.}
  {\bfseries 118} (2017) 021302}
  [\href{https://arxiv.org/abs/1607.07327}{{\ttfamily 1607.07327}}].

\bibitem{Janssen:2014dka}
G.~Janssen et~al., \emph{{Gravitational wave astronomy with the SKA}},
  {\emph{PoS} {\bfseries AASKA14} (2015) 037}
  [\href{https://arxiv.org/abs/1501.00127}{{\ttfamily 1501.00127}}].

\bibitem{Ellis:2023owy}
J.~Ellis, M.~Fairbairn, G.~H\"utsi, M.~Raidal, J.~Urrutia, V.~Vaskonen et~al.,
  \emph{{Prospects for Future Binary Black Hole GW Studies in Light of PTA
  Measurements}},  \href{https://arxiv.org/abs/2301.13854}{{\ttfamily
  2301.13854}}.

\bibitem{KAGRA:2021duu}
{\scshape KAGRA, VIRGO, LIGO Scientific} collaboration, \emph{{Population of
  Merging Compact Binaries Inferred Using Gravitational Waves through GWTC-3}},
  \href{https://doi.org/10.1103/PhysRevX.13.011048}{\emph{Phys. Rev. X}
  {\bfseries 13} (2023) 011048}
  [\href{https://arxiv.org/abs/2111.03634}{{\ttfamily 2111.03634}}].

\bibitem{NANOGrav:2023gor}
{\scshape NANOGrav} collaboration, \emph{{The NANOGrav 15-year Data Set:
  Evidence for a Gravitational-Wave Background}},
  \href{https://doi.org/10.3847/2041-8213/acdac6}{\emph{Astrophys. J. Lett.}
  {\bfseries 951} (2023) } [\href{https://arxiv.org/abs/2306.16213}{{\ttfamily
  2306.16213}}].

\bibitem{Ellis:2023dgf}
J.~Ellis, M.~Fairbairn, G.~H\"utsi, J.~Raidal, J.~Urrutia, V.~Vaskonen et~al.,
  \emph{{Gravitational Waves from SMBH Binaries in Light of the NANOGrav
  15-Year Data}},  \href{https://arxiv.org/abs/2306.17021}{{\ttfamily
  2306.17021}}.

\bibitem{Ellis:2023tsl}
J.~Ellis, M.~Lewicki, C.~Lin and V.~Vaskonen, \emph{{Cosmic Superstrings
  Revisited in Light of NANOGrav 15-Year Data}},
  \href{https://arxiv.org/abs/2306.17147}{{\ttfamily 2306.17147}}.

\bibitem{Sathyaprakash:2012jk}
B.~Sathyaprakash et~al., \emph{{Scientific Objectives of Einstein Telescope}},
  \href{https://doi.org/10.1088/0264-9381/29/12/124013,
  10.1088/0264-9381/30/7/079501}{\emph{Class. Quant. Grav.} {\bfseries 29}
  (2012) 124013} [\href{https://arxiv.org/abs/1206.0331}{{\ttfamily
  1206.0331}}].

\bibitem{LIGOScientific:2021sio}
{\scshape LIGO Scientific, VIRGO, KAGRA} collaboration, \emph{{Tests of General
  Relativity with GWTC-3}},  \href{https://arxiv.org/abs/2112.06861}{{\ttfamily
  2112.06861}}.

\bibitem{Ellis:2020lxl}
J.~Ellis and V.~Vaskonen, \emph{{Probes of gravitational waves with atom
  interferometers}},
  \href{https://doi.org/10.1103/PhysRevD.101.124013}{\emph{Phys. Rev. D}
  {\bfseries 101} (2020) 124013}
  [\href{https://arxiv.org/abs/2003.13480}{{\ttfamily 2003.13480}}].

\bibitem{LIGOScientific:2019fpa}
{\scshape LIGO Scientific, Virgo} collaboration, \emph{{Tests of General
  Relativity with the Binary Black Hole Signals from the LIGO-Virgo Catalog
  GWTC-1}}, \href{https://doi.org/10.1103/PhysRevD.100.104036}{\emph{Phys. Rev.
  D} {\bfseries 100} (2019) 104036}
  [\href{https://arxiv.org/abs/1903.04467}{{\ttfamily 1903.04467}}].

\bibitem{Kiyota:2015dla}
S.~Kiyota and K.~Yamamoto, \emph{{Constraint on modified dispersion relations
  for gravitational waves from gravitational Cherenkov radiation}},
  \href{https://doi.org/10.1103/PhysRevD.92.104036}{\emph{Phys. Rev. D}
  {\bfseries 92} (2015) 104036}
  [\href{https://arxiv.org/abs/1509.00610}{{\ttfamily 1509.00610}}].

\bibitem{Overstreet:2022zgq}
C.~Overstreet, J.~Curti, M.~Kim, P.~Asenbaum, M.A.~Kasevich and F.~Giacomini,
  \emph{{Inference of gravitational field superposition from quantum
  measurements}},  \href{https://arxiv.org/abs/arXiv:2209.02214}{{\ttfamily
  arXiv:2209.02214}}.

\end{thebibliography}\endgroup

\end{document}